\documentstyle[preprint,aps,eqsecnum]{revtex}
\begin{document}
\draft
\tightenlines

\newcommand{\beq}{\begin{equation}}
\newcommand{\eeq}{\end{equation}}
\newcommand{\bea}{\begin{eqnarray}}
\newcommand{\eea}{\end{eqnarray}}
\newcommand{\cir}{{\buildrel \circ \over =}}

\title{ Multipolar Expansions for the  Relativistic
 N-Body Problem in the Rest-Frame Instant Form.}

\author{David Alba}

\address
{Dipartimento di Fisica\\
Universita' di Firenze\\
L.go E.Fermi 2 (Arcetri)\\
50125 Firenze, Italy\\
E-mail: ALBA@FI.INFN.IT}

\author{and}

\author{Luca Lusanna}

\address
{Sezione INFN di Firenze\\
L.go E.Fermi 2 (Arcetri)\\
50125 Firenze, Italy\\
E-mail: LUSANNA@FI.INFN.IT}

\author{and}

\author{Massimo Pauri}

\address
{Dipartimento di Fisica\\ Universita' di Parma\\
 Parco Area Scienze 7/A\\
  43100 Parma, Italy\\
   E-mail: PAURI@PR.INFN.IT}

\maketitle

\begin{abstract}

Dixon's multipoles for a system of N relativistic positive-energy
scalar particles are evaluated in the rest-frame instant form of
dynamics. The Wigner hyperplanes (intrinsic rest frame of the
isolated system) turn out to be the natural framework for
describing multipole kinematics. In particular, concepts like the
{\it barycentric tensor of inertia} can be defined in special
relativity only by means of the quadrupole moments of the isolated
system.

\vskip 1truecm

\today

\vskip 1truecm

\end{abstract}
\vfill\eject

\vfill\eject

\section{Introduction.}

In a recent paper\cite{iten1} we have given a complete treatment
of the kinematics of the relativistic N-body problem in the {\it
rest-frame instant form of dynamics}
\cite{lus,albad,crater,india}. We have shown in particular how to
perform the separation of the center-of-mass motion in the
relativistic case. This requires the reformulation of the theory
of relativistic isolated systems on arbitrary spacelike
hypersurfaces \footnote{Such hypersurfaces are the leaves of a
foliation of Minkowski spacetime (namely one among its 3+1
splittings) and are equivalent to a congruence of timelike
accelerated observers.}. This description is also able to
incorporate the coupling to the gravitational field. This is
essentially  Dirac's reformulation\cite{dirac} of classical field
theory (suitably extended to particles) on arbitrary spacelike
hypersurfaces ({\it equal time} surfaces) and provides the
classical basis of the Tomonaga-Schwinger formulation of quantum
field theory. For each isolated system (containing any combination
of particles, strings and fields) one gets a reformulation as a
parametrized Minkowski theory\cite{lus}, with the extra bonus of
having the theory already predisposed to the coupling to gravity
in its ADM formulation, but with the price that the functions
$z^{\mu}(\tau ,\vec \sigma )$ describing the embedding of the
spacelike hypersurface in Minkowski spacetime become configuration
variables in the action principle. Since the action is invariant
under separate $\tau$-reparametrizations and
space-diffeomorphisms, there emerge first class constraints
ensuring the independence of the description from the choice of
the 3+1 splitting. The embedding configuration variables
$z^{\mu}(\tau ,\vec \sigma )$ are the {\it gauge} variables
associated with this kind of general covariance.

Let us remark that, since the intersection of a timelike worldline
with a spacelike hypersurface corresponding to a value $\tau$ of
the time parameter is identified by 3 numbers $\vec \sigma =\vec
\eta (\tau )$ instead of four, in parametrized Minkowski theories
each particle must have a well defined sign of the energy:
therefore we cannot describe the two topologically disjoint
branches of the mass hyperboloid simultaneously like in the
standard manifestly Lorentz-covariant theory. As a consequence
there are no more mass-shell constraints. Therefore, each particle
with a definite sign of the energy is described by the canonical
coordinates ${\vec \eta}_i(\tau )$, ${\vec \kappa}_i(\tau )$ with
the derived 4-position of the particles given by $x^{\mu}_i(\tau
)=z^{\mu}(\tau ,{\vec \eta}_i(\tau ))$. The derived 4-momenta
$p^{\mu}_i(\tau )$ are ${\vec \kappa}_i$-dependent solutions of
$p^2_i-\epsilon m^2_i =0$ with the chosen sign of the energy.

As said, parametrized Minkowski theories have separate spatial and
time reparametrization invariances, which imply the independence of
the description from the choice of the 3+1 splitting of Minkowski
spacetime. In Minkowski spacetime we can restrict the foliation to
have spacelike {\it hyperplanes} as leaves. Then, for each
configuration of the isolated system with timelike 4-momentum, we can
restrict ourselves to the special foliation whose leaves are the
hyperplanes orthogonal to the conserved system 4-momentum, which have
been named {\it Wigner hyperplanes}. This special foliation is
intrinsically determined only by the configuration of the isolated
system. In this way\cite{lus} it is possible to  arrive at the
definition of the {\it Wigner-covariant rest-frame instant form of
dynamics} for every isolated system whose configurations have well
defined and finite Poincar\'e generators with timelike total
4-momentum (see Ref.\cite{dir} for the traditional forms of dynamics).

This formulation  clarifies the non-trivial definition of a
relativistic center of mass. As well known, no such definition can
enjoy all the properties of the non-relativistic center of mass.
See Refs.\cite{pau,com1,com2,com3,com4} for a partial bibliography
of all the existing attempts.

As shown in Appendix A of Ref.\cite{iten1}   only four first class
constraints survive in the rest-frame instant form on Wigner
hyperplanes and the original configuration variables $z^{\mu}(\tau
,\vec \sigma )$, ${\vec \eta}_i(\tau )$ and their conjugate
momenta $\rho_{\mu}(\tau ,\vec \sigma )$, ${\vec \kappa}_i(\tau )$
are reduced to:

i) a decoupled particle ${\tilde x}^{\mu}_s(\tau )$, $p^{\mu}_s$
(the only remnant of the spacelike hypersurface) with a positive
mass $\epsilon_s=\sqrt{\epsilon p^2_s}$ determined by the first
class constraint $\epsilon_s-M_{sys} \approx 0$
\footnote{$M_{sys}$ being the invariant mass of the isolated
system.} and with its rest-frame Lorentz scalar time
$T_s={{{\tilde x}_s\cdot p_s}\over {\epsilon_s}}$ put equal to the
mathematical time as the gauge fixing $T_s-\tau \approx 0$ to the
previous constraint. Here, ${\tilde x}^{\mu}_s(\tau )$ is a {\it
non-covariant canonical} variable for the {\it 4-center of mass}.
After the elimination of $T_s$ and $\epsilon_s$ with the previous
pair of second class constraints, one remains with a decoupled
free point ({\it point particle clock}) of mass $M_{sys}$ and
canonical 3-coordinates ${\vec z}_s=\epsilon_s ({\vec {\tilde
x}}_s-{{{\vec p}_s}\over {p^o_s}} {\tilde x}^o)$, ${\vec
k}_s={{{\vec p}_s}\over {\epsilon_s}}$ \footnote{${\vec
z}_s/\epsilon_s$ is the classical analogue of the Newton-Wigner
3-position operator, only covariant under the Euclidean subgroup
of the Poincar\'e group.}. The non-covariant canonical ${\tilde
x}^{\mu}_s(\tau )$ must not be confused with the 4-vector
$x^{\mu}_s(\tau )=z^{\mu}(\tau ,\vec \sigma =0)$ identifying the
origin of the 3-coordinates $\vec \sigma$ inside the Wigner
hyperplanes. The worldline $x^{\mu}_s(\tau )$ is arbitrary because
it depends on $x^{\mu}_s(0)$ and its 4-velocity ${\dot
x}^{\mu}_s(\tau )$ depends on the Dirac multipliers associated
with the 4 left first class constraints \footnote{Therefore this
arbitrary worldline may be considered as an arbitrary {\it
centroid} for the isolated system.}, as it will be shown in the
next Section. The unit timelike 4-vector
$u^{\mu}(p_s)=p_s^{\mu}/\epsilon_s$ is orthogonal to the Wigner
hyperplanes and describes their orientation in the chosen inertial
frame.

ii) the particle canonical variables ${\vec \eta}_i(\tau )$, ${\vec
\kappa}_i(\tau )$ \footnote{They are Wigner spin-1 3-vectors, like the coordinates
$\vec \sigma$.} inside the Wigner hyperplanes. They are restricted by
the three first class constraints (the {\it rest-frame conditions})
${\vec \kappa}_{+}=\sum_{i=1}^N {\vec \kappa}_i \approx 0$. Since the
role of the relativistic decoupled {\it 4-center of mass} is taken by
${\tilde x}^{\mu}_s(\tau )$ (or,after the gauge fixing $T_s-\tau
\approx 0$, by an {\it external 3-center of mass} ${\vec z}_s$,
defined in terms of ${\tilde x}^{\mu}_s$ and $p^{\mu}_s$
\cite{iten1}), the rest-frame conditions imply that the {\it internal}
3-center of mass ${\vec q}_{+}={\vec
\sigma}_{com}$ is a {\it gauge variable}, which can be eliminated with gauge fixings
\footnote{For instance ${\vec q}_{+}\approx 0$ implies that the {\it internal 3-center
of mass} is put in the origin $x^{\mu}_s(\tau )=z^{\mu}(\tau ,\vec
\sigma =0)$.}.

Therefore, we need a doubling of the concepts:

1) There is the {\it external} viewpoint of an arbitrary inertial
Lorentz observer, who describes the Wigner hyperplanes, leaves of a
foliation of Minkowski spacetime, determined by the timelike
configurations of the isolated system. A change of inertial observer
by means of a Lorentz transformation rotates the Wigner hyperplanes
and induces a Wigner rotation of the 3-vectors inside each Wigner
hyperplane. Every such hyperplane inherits an induced {\it internal
Euclidean structure} while an {\it external} realization of the
Poincar\'e group induces the {\it internal} Euclidean action. As said
above, an arbitrary worldline  ({\it centroid}) $x^{\mu}_s(\tau )$ is
chosen as origin of the {\it internal} 3-coordinates on the Wigner
hyperplanes.

Three {\it external} concepts of 4-center of mass can be defined by
using the {\it external} realization of the Poincar\'e algebra (to
each one there corresponds a 3-location inside the Wigner
hyperplanes):

a) the {\it external} non-covariant canonical {\it 4-center of
mass} (also named {\it 4-center of spin}) ${\tilde x}^{\mu}_s$
(with 3-location ${\vec {\tilde \sigma}}$),

b) the {\it external} non-covariant non-canonical M\o ller {\it
4-center of energy} $R^{\mu}_s$ (with 3-location ${\vec
\sigma}_R$),

c) the {\it external} covariant non-canonical Fokker-Pryce {\it
4-center of inertia} $Y^{\mu}_s$ (with 3-location ${\vec
\sigma}_Y$).

Only the canonical non-covariant center of mass ${\tilde
x}^{\mu}_s(\tau )$ is relevant in the Hamiltonian treatment with
Dirac constraints, while only the Fokker-Pryce $Y^{\mu}_s$ is a
4-vector by construction. See Ref.\cite{iten1} for the
construction of the {\it 4-centers} starting from the
corresponding {\it 3-centers} (3-center of spin\cite{com3},
3-center of energy \cite{mol}, 3-center of
inertia\cite{com2,com3}).

2) There is the {\it internal} viewpoint inside the Wigner hyperplanes
associated to a unfaithful {\it internal} realization of the
Poincar\'e algebra: the total {\it internal} 3-momentum of the
isolated system vanishes due to the rest-frame conditions. The {\it
internal} energy and angular momentum are the invariant mass $M_{sys}$
and the spin (the angular momentum with respect to ${\tilde
x}^{\mu}_s(\tau )$) of the isolated system respectively. With the {\it
internal} realization of the Poincar\'e algebra we can define three
{\it internal} 3-centers of mass: the {\it internal} canonical
3-center of mass, the {\it internal} M\o ller 3-center of energy and
the {\it internal} Fokker-Pryce 3-center of inertia. But, due to the
rest-frame conditions, {\it they coincide}. As a natural gauge fixing
to the rest-frame conditions we can add the vanishing of the {\it
internal} Lorentz boosts: it is equivalent to locate the internal
canonical 3-center of mass ${\vec q}_{+}$ in $\vec
\sigma =0$, i.e. in the origin $x^{\mu}_s(\tau )=z^{\mu}(\tau ,\vec 0)$. With
these gauge fixings and with $T_s-\tau \approx 0$, the worldline
$x^{\mu}_s(\tau )$ becomes uniquely determined except for the
arbitrariness in the choice of $x^{\mu}_s(0)$ [$\,
u^{\mu}(p_s)=p^{\mu}_s/\epsilon_s$]

\beq
x^{\mu}_s(\tau )=x^{\mu}_s(0) + u^{\mu}(p_s) T_s,
\label{I1}
\eeq

\noindent   and coincides with the {\it
external} covariant non-canonical Fokker-Pryce 4-center of inertia,
$x^{\mu}_s(\tau ) = x^{\mu}_s(0) + Y^{\mu}_s$\cite{iten1}.

This doubling of concepts with the {\it external} non-covariant
canonical 4-center of mass ${\tilde x}^{\mu}_s(\tau )$ (or with the
{\it external} 3-center of mass ${\vec z}_s$ when $T_s-\tau \approx
0$) and with the {\it internal} canonical 3-center of mass ${\vec
q}_{+}\approx 0$ {\it replaces} the separation of the non-relativistic
3-center of mass due to the Abelian translation symmetry. The
non-relativistic conserved 3-momentum is replaced by the {\it
external} ${\vec p}_s=\epsilon_s {\vec k}_s$, while the {\it internal}
3-momentum vanishes, ${\vec \kappa}_{+}\approx 0$, as a definition of
rest frame.

In the final gauge we have $\epsilon_s \equiv M_{sys}$, $T_s
\equiv \tau $ and the canonical basis ${\vec z}_s$, ${\vec k}_s$,
${\vec \eta}_i$, ${\vec \kappa}_i$ is restricted by the three
pairs of second class constraints ${\vec
\kappa}_{+}=\sum_{i=1}^N{\vec \kappa}_i \approx 0$, ${\vec q}_{+}
\approx 0$, so that 6N canonical variables describe the N
particles like in the non-relativistic case. We still need a
canonical transformation ${\vec \eta}_i$, ${\vec \kappa}_i$ $\,\,
\mapsto \,\,$ ${\vec q}_{+} [\approx 0]$, ${\vec \kappa}_{+}
[\approx 0]$, ${\vec \rho}_a$, ${\vec \pi}_a$ [$a=1,..,N-1$]
identifying a set of relative canonical variables. The final
6N-dimensional canonical basis is ${\vec z}_s$, ${\vec k}_s$,
${\vec \rho}_a$, ${\vec \pi}_a$. To get this result  we need a
highly non-linear canonical transformation\cite{iten1}, which can
be obtained by exploiting the Gartenhaus-Schwartz singular
transformation \cite{garten}.

In the end we obtain the Hamiltonian for relative motions as a sum of
N square roots, each one containing a squared mass and a quadratic
form in the relative momenta, which goes into the non-relativistic
Hamiltonian for relative motions in the limit $c\,
\rightarrow \infty$. This fact has the following implications:

a) if one tries to make the inverse Legendre transformation to
find the associated Lagrangian, it turns out that, due to  the
presence of square roots, the Lagrangian is a hyperelliptic
function of ${\dot {\vec \rho}}_a$ already in the free case. A
closed form exists only for N=2, $m_1=m_2=m$: $L=-\epsilon m
\sqrt{4-{\dot {\vec \rho}}^2}$. This exceptional case already
shows that the existence of the limiting velocity  $c$ (or in
other terms the Lorentz signature of spacetime) forbids a linear
relation between the spin (center-of-mass angular momentum) and
the angular velocity.

b) the N quadratic forms in the relative momenta appearing in the
relative Hamiltonian cannot be simultaneously diagonalized. In any
case the Hamiltonian is a sum of square roots, so that concepts
like {\it reduced masses}, {\it Jacobi normal relative
coordinates} and {\it tensor of inertia} cannot be extended to
special relativity. As a consequence, for example, a relativistic
static orientation-shape SO(3) principal bundle approach
\cite{little}\footnote{See Ref.\cite{iten1} for a review of this
approach used in molecular physics for the definition and study of
the {\it vibrations} of molecules.} can be implemented only by
using non-Jacobi relative coordinates.

c) the best way of studying rotational kinematics (the non-Abelian
rotational symmetry, associated with the conserved {\it internal}
spin) is based on the {\it canonical spin bases} with the
associated concepts of {\it spin frames} and {\it dynamical body
frames} introduced in Ref.\cite{iten1}: they can be build in the
same way as in the non-relativistic case \cite{iten2} starting
from the canonical basis ${\vec \rho}_a$, ${\vec \pi}_a$.

Let us clarify this point.

In the non-relativistic N-body problem it is easy to make the
separation of the absolute translational motion of the center of mass
from the relative motions, due to the Abelian nature of the
translation symmetry group. This implies that the associated Noether
constants of motion (the conserved total 3-momentum) are in
involution, so that the center-of-mass degrees of  freedom decouple.
Moreover, the fact that the non-relativistic kinetic energy of the
relative motions is a quadratic form in the relative velocities allows
the introduction of special sets of relative coordinates, the {\it
Jacobi normal relative coordinates}, which diagonalize the quadratic
form and correspond to different patterns of clustering of the centers
of mass of the particles. Each set of Jacobi normal relative
coordinates organizes the N particles into a {\it hierarchy of
clusters}, in which each cluster of two or more particles has a mass
given by an eigenvalue ({\it reduced masses}) of the quadratic form;
Jacobi normal coordinates join the centers of mass of pairs of
clusters.

However, the non-Abelian nature of the rotation symmetry group,
whose associated Noether constants of motion (the conserved total
angular momentum) are not in involution, prevents the possibility
of a global separation of absolute rotations from the relative
motions, so that there is no global definition of absolute {\it
vibrations}. This has the consequence that an {\it isolated}
deformable body can undergo rotations by changing its own shape
(see the examples of the {\it falling cat} and of the {\it
diver}). It was just to deal with these problems that the theory
of the orientation-shape SO(3) principal bundle
approach\cite{little} has been developed. Its essential content is
that any {\it static} (i.e. velocity-independent) definition of
{\it body frame} for a deformable body must be interpreted as a
gauge fixing in the context of a SO(3) {\it gauge} theory. Both
the laboratory and the body frame angular velocities as well as
the orientational variables of the static body frame become
thereby {\it unobservable gauge} variables. This approach is
associated with a set of {\it point} canonical transformations,
which allow to define the body frame components of relative
motions in a velocity-independent way.

Since in many physical applications (e.g. nuclear physics,
rotating stars,...) angular velocities are viewed as {\it
measurable} quantities, it is desiderable to have an alternative
formulation complying with this requirement, possibly
generalizable to special relativity. This program has been
realized in a previous paper \cite{iten2}, of which
Ref.\cite{iten1} is the relativistic extension. Let us summarize
the main points of our formulation.

First of all for $N \geq 3$ we have constructed in
Ref.\cite{iten2} a class of {\it non-point} canonical
transformations, which allows to build the {\it canonical spin
bases} quoted above, which are connected to the patterns of the
possible {\it clusterings of the spins} associated with relative
motions. The definition of these {\it spin bases} is independent
of Jacobi normal relative coordinates, just as the patterns of
spin clustering are independent of the patterns of center-of-mass
Jacobi clustering. We have  found two basic frames associated to
each spin basis: the {\it spin frame} and the {\it dynamical body
frame}. Their construction is  guaranteed by the fact that,
besides the existence on the relative phase space  of a
Hamiltonian symmetry {\it left} action of SO(3)\footnote{We adhere
to the definitions used in Ref.\cite{little}; in the mathematical
literature our {\it left} action is a {\it right} action.}
\footnote{Their generators are the center-of-mass angular momentum
Noether constants of motion.},  it is possible to define as many
Hamiltonian non-symmetry  {\it right} actions of SO(3)
\footnote{Their generators are not constants of motion.} as the
possible patterns of spin clustering. While for N=3 the unique
canonical spin basis coincides with a special class of global
cross sections of the trivial orientation-shape SO(3) principal
bundle, for $N \geq 4$ the existing {\it spin bases} and {\it
dynamical body frames} turn out to be unrelated  to the local
cross sections of the {\it static} non-trivial orientation-shape
SO(3) principal bundle, and {\it evolve} in a dynamical way
dictated by the equations of motion. In this new formulation {\it
both} the orientation variables and the angular velocities become
{\it measurable} quantities in each canonical spin basis by
construction.

For each N every allowed spin basis provides a physically
well-defined separation between {\it rotational} and {\it
vibrational} degrees of freedom. The non-Abelian nature of the
rotational symmetry implies that there is no unique separation of
{\it absolute rotations} and {\it relative motions}. The unique
{\it body frame} of rigid bodies is replaced by a discrete number
of {\it evolving dynamical body frames} and of {\it spin canonical
bases}, both of which are grounded on patterns of spin couplings,
direct analogues of the coupling  of quantum angular momenta.

In this paper we complete our study of relativistic kinematics for
the N-body system by evaluating its rest-frame Dixon multipoles
\cite{dixon}\footnote{See Ref.\cite{dixon1} for the definition of
Dixon's multipoles in general relativity.}. Let us recall that
this method is being used  for treating extended systems in
astrophysics. The starting point is the definition of the energy
momentum tensor of the N positive-energy particles on the Wigner
hyperplane. In this way we see that the Wigner hyperplane is the
natural framework for reorganizing a lot of kinematics connected
with multipoles. Moreover, only in this way, a concept like the
{\it barycentric tensor of inertia} can be introduced in special
relativity by means of the quadrupole moments.

A review of the rest-frame instant form of dynamics for N scalar free
positive-energy particles is given in Section II.

In Section III we evaluate the energy momentum tensor on the Wigner
hyperplanes.

Dixon's multipoles for the system are defined in Section IV. A special
study of monopole, dipole and quadrupole moments is given together
with the multipolar expansion.

Other properties of Dixon's multipoles are reviewed in Section V.

Some comments on open problems are given in the Conclusions.

The non-relativistic N-particle multipolar expansion is given in
Appendix A, while in Appendix B there is a review of symmetric
trace-free  (STF) tensors.

\vfill\eject

\section{Review of the Rest-Frame Instant Form.}

Let us review the  system of N free scalar positive-energy
particles in the framework of parametrized Minkowski theory (see
Appendices A and B and Section II  of Ref.\cite{iten1}). As said
in the Introduction, each particle is described by a configuration
3-vector ${\vec \eta}_i(\tau )$. The particle worldline is $x^{\mu
}_i(\tau )=z^{\mu }(\tau ,{\vec \eta}_i(\tau ))$, where
$z^{\mu}(\tau ,\vec \sigma )$ are the embedding configuration
variables descibing the spacelike hypersurface $\Sigma_{\tau}$
\footnote{The foliation is defined by an embedding $R\times \Sigma
\rightarrow M^4$, $(\tau ,\vec \sigma ) \mapsto z^{\mu}(\tau ,\vec
\sigma )$, with $\Sigma$ an abstract 3-surface diffeomorphic to
$R^3$. $\Sigma_{\tau}$ is the Cauchy surface of {\it equal time}.
The metric induced on it is $g_{ AB}[z]
=z^{\mu}_{A}\eta_{\mu\nu}z^{\nu}_{B}$ on $\Sigma_{\tau}$, a
functional of $z^{\mu}$, and the embedding coordinates
$z^{\mu}(\tau ,\vec \sigma )$ are considered as independent
fields. We use the notation $\sigma^{A}=(\tau ,\sigma^{\check r})$
of Refs.\cite{lus,crater}. The $z^{\mu}_{A}(\sigma )=
\partial z^{\mu}(\sigma )/\partial \sigma^{A}$ are flat cotetrad fields on Minkowski
spacetime with the $z^{\mu}_{\check r}$'s tangent to $\Sigma_{\tau}$.
While in Ref.\cite{iten1} we used the metric convention
$\eta_{\mu\nu}=\epsilon (+---)$ with $\epsilon =\pm$, in this paper we
shall use $\epsilon =1$ like in Ref.\cite{lus}.}.

The system is described by the action\cite{lus,albad,crater}

\begin{eqnarray}
S&=& \int d\tau d^3\sigma \, {\cal L}(\tau ,\vec
\sigma )=\int d\tau L(\tau ),\nonumber \\
 &&{\cal L}(\tau ,\vec \sigma )=-\sum_{i=1}^N\delta^3(\vec \sigma -{\vec \eta}_i
(\tau ))m_i\sqrt{ g_{\tau\tau}(\tau ,\vec \sigma )+2g_{\tau {\check
r}} (\tau ,\vec \sigma ){\dot \eta}^{\check r}_i(\tau )+g_{{\check
r}{\check s}} (\tau ,\vec \sigma ){\dot \eta}_i^{\check r}(\tau ){\dot
\eta}_i^{\check s} (\tau )  },\nonumber \\
 &&L(\tau
)=-\sum_{i=1}^Nm_i\sqrt{ g_{\tau\tau}(\tau ,{\vec \eta}_i (\tau
))+2g_{\tau {\check r}}(\tau ,{\vec \eta}_i(\tau )){\dot \eta}^{\check
r}
_i(\tau )+g_{{\check r}{\check s}}(\tau ,{\vec \eta}_i(\tau )){\dot \eta}_i
^{\check r}(\tau ){\dot \eta}_i^{\check s}(\tau )  }.
\label{II1}
\end{eqnarray}

\noindent The action is invariant under separate $\tau$- and $\vec
\sigma$-reparametrizations.

The canonical momenta are

\begin{eqnarray}
\rho_{\mu}(\tau ,\vec \sigma )&=&-{ {\partial {\cal L}(\tau ,\vec \sigma )}
\over {\partial z^{\mu}_{\tau}(\tau ,\vec \sigma )} }=\sum_{i=1}^N\delta^3
(\vec \sigma -{\vec \eta}_i(\tau ))m_i\nonumber \\
 &&{{z_{\tau\mu}(\tau ,\vec \sigma )+z_{{\check r}\mu}(\tau ,\vec \sigma )
{\dot \eta}_i^{\check r}(\tau )}\over {\sqrt{g_{\tau\tau}(\tau ,\vec
\sigma )+ 2g_{\tau {\check r}}(\tau ,\vec \sigma ){\dot
\eta}_i^{\check r}(\tau )+ g_{{\check r}{\check s}}(\tau ,\vec \sigma
){\dot \eta}_i^{\check r}(\tau ){\dot
\eta}_i^{\check s}(\tau ) }} }=\nonumber \\
&=&[(\rho_{\nu}l^{\nu})l_{\mu}+(\rho_{\nu}z^{\nu}_{\check r})\gamma^{{\check r}
{\check s}}z_{{\check s}\mu}](\tau ,\vec \sigma ),\nonumber \\
 &&{}\nonumber \\
\kappa_{i{\check r}}(\tau )&=&-{ {\partial L(\tau )}\over {\partial {\dot
\eta}_i^{\check r}(\tau )} }=\nonumber \\
&=&m_i{ {g_{\tau {\check r}}(\tau ,{\vec \eta}_i(\tau ))+g_{{\check r}
{\check s}}(\tau ,{\vec \eta}_i(\tau )){\dot
\eta}_i^{\check s}(\tau )}\over { \sqrt{g_{\tau\tau}(\tau ,{\vec
\eta}_i(\tau ))+ 2g_{\tau {\check r}}(\tau ,{\vec \eta}_i(\tau )){\dot
\eta}_i^{\check r}(\tau )+ g_{{\check r}{\check s}}(\tau ,{\vec
\eta}_i(\tau )){\dot \eta}_i^{\check r} (\tau ){\dot \eta}_i^{\check
s}(\tau ) }} },\nonumber \\
 &&{}\nonumber \\
&&\lbrace z^{\mu}(\tau ,\vec \sigma ),\rho_{\nu}(\tau ,{\vec
\sigma}^{'}\rbrace
=-\eta^{\mu}_{\nu}\delta^3(\vec \sigma -{\vec \sigma}^{'}),\nonumber \\
&&\lbrace \eta^{\check r}_i(\tau ),\kappa_{j{\check s}}(\tau )\rbrace =
-\delta_{ij}\delta^{\check r}_{\check s}.
\label{II2}
\end{eqnarray}

The canonical Hamiltonian $H_{c}$ is zero, but there are the primary
first class constraints

\begin{eqnarray}
{\cal H}_{\mu}(\tau ,\vec \sigma )&=& \rho_{\mu}(\tau ,\vec \sigma )-l_{\mu}
(\tau ,\vec \sigma )\sum_{i=1}^N\delta^3(\vec \sigma -{\vec \eta}_i(\tau ))
\sqrt{ m^2_i-\gamma^{{\check r}{\check s}}(\tau ,\vec \sigma )
\kappa_{i{\check r}}(\tau )\kappa_{i{\check s}}(\tau ) }-\nonumber \\
&-&z_{{\check r}\mu}
(\tau ,\vec \sigma )\gamma^{{\check r}{\check s}}(\tau ,\vec \sigma )
\sum_{i=1}^N\delta^3(\vec \sigma -{\vec \eta}_i(\tau ))\kappa_{i{\check s}}
\approx 0,
\label{II3}
\end{eqnarray}

\noindent so that the Dirac Hamiltonian is $H_D=\int d^3\sigma
\lambda^{\mu}(\tau ,\vec \sigma ) {\cal H}_{\mu}(\tau ,\vec \sigma
)$, with the $\lambda^{\mu}(\tau ,\vec \sigma )$ Dirac
multipliers.

The conserved Poincar\'e generators are (the suffix ``s" denotes the
hypersurface $\Sigma_{\tau}$)

\begin{eqnarray}
&&p^{\mu}_s=\int d^3\sigma \rho^{\mu}(\tau ,\vec \sigma ),\nonumber \\
&&J_s^{\mu\nu}=\int d^3\sigma [z^{\mu}(\tau ,\vec \sigma )\rho^{\nu}(\tau ,
\vec \sigma )-z^{\nu}(\tau ,\vec \sigma )\rho^{\mu}(\tau ,\vec \sigma )].
\label{II4}
\end{eqnarray}

After the restriction to spacelike {\it hyperplanes} the Dirac
Hamiltonian is reduced to $H_D=\lambda_{\mu}(\tau ){\tilde {\cal
H}}^{\mu}(\tau )+\lambda_{\mu\nu}(\tau ){\tilde {\cal
H}}^{\mu\nu}(\tau )$  (only ten Dirac multipliers survive) with
the remaining ten constraints given by

\begin{eqnarray}
{\tilde {\cal H}}^{\mu}(\tau )&=&\int d^3\sigma {\cal H}^{\mu}(\tau ,\vec
\sigma )=\, p^{\mu}_s-l^{\mu}\sum_{i=1}^N\sqrt{m^2_i+{\vec \kappa}^2_i
(\tau )}+b^{\mu}_{\check r}(\tau )\sum_{i=1}^N\kappa_{i{\check r}}(\tau )
\approx 0,\nonumber \\
{\tilde {\cal H}}^{\mu\nu}(\tau )&=&b^{\mu}_{\check r}(\tau )\int d^3\sigma
\sigma^{\check r}\, {\cal H}^{\nu}(\tau ,\vec \sigma )-b^{\nu}_{\check r}(\tau )
\int d^3\sigma \sigma^{\check r}\, {\cal H}^{\mu}(\tau ,\vec \sigma )=
\nonumber \\
&=&S_s^{\mu\nu}(\tau )-[b^{\mu}_{\check r}(\tau
)b^{\nu}_{\tau}-b^{\nu}_{\check r}(\tau
)b^{\mu}_{\tau}]\sum_{i=1}^N\eta_i^{\check r}(\tau )
\sqrt{m^2_i+{\vec \kappa}^2_i(\tau )}-\nonumber \\
&-&[b^{\mu}_{\check r}(\tau )b^{\nu}_{\check s}(\tau )-b^{\nu}_{\check r}(\tau )
b^{\mu}_{\check s}(\tau )]\sum_{i=1}^N\eta_i^{\check r}(\tau )\kappa_i^{\check
s}(\tau )\approx 0.
\label{II5}
\end{eqnarray}

Here $S^{\mu\nu}_s$ is the spin part of the Lorentz generators

\begin{eqnarray}
J^{\mu\nu}_s&=&x^{\mu}_sp^{\nu}_s-x^{\nu}_sp^{\mu}_s+S^{\mu\nu}_s,\nonumber \\
&&S^{\mu\nu}_s=b^{\mu}_{\check r}(\tau )\int d^3\sigma \sigma^{\check r}
\rho^{\nu}(\tau ,\vec \sigma )-b^{\nu}_{\check r}(\tau )\int d^3\sigma
\sigma^{\check r}\rho^{\mu}(\tau ,\vec \sigma ).
\label{II6}
\end{eqnarray}

On the {\it Wigner hyperplane}\footnote{On it we use the notation
$A=(\tau ,r)$. The 3-vectors $\vec B=\{ b^r\}$ on the Wigner
hyperplanes are Wigner spin-1 3-vectors.} we have the following
constraints and Dirac Hamiltonian\cite{lus,crater}
\footnote{$\epsilon^{\mu}_r(u(p_s))=L^{\mu}{}_r(p_s, {\buildrel
\circ \over p}_s)$ and
$\epsilon^{\mu}_{\tau}(u(p_s))=u^{\mu}(p_s)=L^{\mu}{}_o(p_s,{\buildrel
\circ \over p}_s)$ are the columns of the standard Wigner boost
for timelike Poincar\'e orbits. See Appendix B of
Ref.\cite{iten1}.}

\begin{eqnarray}
{\tilde {\cal H}}^{\mu}(\tau )&=&p^{\mu}_s-u^{\mu}(p_s) \sum_{i=1}^N
\sqrt{m_i
^2+{\vec \kappa}_i^2} +\epsilon^{\mu}_r(u(p_s)) \sum_{i=1}^N\kappa_{ir}=
\nonumber \\
&=&u^{\mu}(p_s) [\epsilon_s-\sum_{i=1}^N\sqrt{m_i^2+ {\vec
\kappa}_i^2}] +\epsilon^{\mu}_r(u(p_s)) \sum_{i=1}^N \kappa_{ir}
\approx 0,\nonumber \\
&&{}\nonumber \\ &&or\nonumber \\
 &&{}\nonumber \\
 &&\epsilon_s-M_{sys}\approx 0,\quad\quad
M_{sys}=\sum_{i=1}^N\sqrt{m_i^2+{\vec \kappa}_i^2},\nonumber \\
 &&{\vec p}_{sys} = {\vec
\kappa}_{+}=\sum_{i=1}^N {\vec \kappa}_i \approx 0,\nonumber \\
 &&{}\nonumber \\
 H_D&=&
\lambda^{\mu}(\tau ) {\tilde {\cal H}}_{\mu}(\tau )=
\lambda (\tau ) [\epsilon_s-M_{sys}]-\vec \lambda (\tau ) \sum_{i=1}^N {\vec \kappa}_i,
\nonumber \\
&&{}\nonumber \\
&&\lambda (\tau ) \approx -{\dot x}_{s \mu}(\tau )u^{\mu}(p_s),\nonumber \\
&&\lambda_r(\tau )\approx -{\dot x}_{s \mu}(\tau )\epsilon^{\mu}_r(u(p_s)),
\nonumber \\
&&{}\nonumber \\
 {\dot {\tilde x}}^{\mu}_s(\tau ) &=&-\lambda (\tau )
u^{\mu}(p_s),\nonumber \\
 &&{}\nonumber \\
 {\dot x}_s^{\mu}(\tau )  &{\buildrel \circ \over =}\,& \{
x^{\mu}_s(\tau ), H_D \} ={\tilde \lambda}_{\nu}(\tau ) \{
x^{\mu}_s(\tau ), {\cal H}^{\mu}(\tau ) \} \approx\nonumber \\
 &\approx& -{\tilde \lambda}^{\mu}(\tau )=-\lambda (\tau )
u^{\mu}(p_s)+\epsilon^{\mu}_r(u(p_s)) \lambda_r(\tau ),\nonumber \\
 &&{}\nonumber \\
{\dot x}^2_s(\tau )&=& \lambda^2(\tau )-{\vec \lambda}^2(\tau ) >
0,\quad\quad {\dot x}_s\cdot u(p_s)=-\lambda (\tau ),\nonumber \\
 &&{}\nonumber \\
U^{\mu}_s(\tau )&=& {{ {\dot x}^{\mu}_s(\tau )}\over { \sqrt{{\dot
x}^2
_s(\tau )} }}={{-\lambda (\tau )u^{\mu}(p_s)+\lambda_r(\tau )\epsilon^{\mu}_r
(u(p_s))}\over {\sqrt{\lambda^2(\tau )-{\vec \lambda}^2(\tau )} }},\nonumber \\
&&{}\nonumber \\
\Rightarrow&& x^{\mu}_s(\tau )=x^{\mu}_s(0)-u^{\mu}(p_s)\int_0^{\tau}d\tau_1
\lambda (\tau_1)+\epsilon^{\mu}_r(u(p_s))\int_0^{\tau}d\tau_1\lambda_r(\tau_1).
\label{II7}
\end{eqnarray}

While the Dirac multiplier $\lambda (\tau )$  is determined by the
gauge fixing $T_s-\tau \approx 0$, the 3 Dirac's multipliers $\vec
\lambda (\tau )$ describe the classical zitterbewegung of the
origin $x^{\mu}_s(\tau )=z^{\mu}(\tau ,\vec 0)$ of the coordinates
on the Wigner hyperplane. Each gauge-fixing $\vec \chi (\tau
)\approx 0$ to the 3 first class constraints ${\vec
\kappa}_{+}\approx 0$ (defining the {\it internal rest-frame})
gives a different determination of the multipliers $\vec \lambda
(\tau )$. Therefore it identifies a different worldline for the
covariant non-canonical origin $x^{(\vec \chi )\mu}_s(\tau )$
which carries with itself the definition ${\vec \sigma}_{com}$ of
the {\it internal 3-center of mass}, conjugate with ${\vec
\kappa}_{+}$ \footnote{Obviously each choice $\vec \chi (\tau )$
leads to a different set of conjugate canonical relative
variables.}.

 Let us remark that the
constant $x^{\mu}_s(0)$ [and, therefore, also ${\tilde x}
^{\mu}_s(0)$] is arbitrary, reflecting the arbitrariness in the
absolute location of the origin of the {\it internal} coordinates
on each hyperplane in Minkowski spacetime. The origin
$x^{\mu}_s(\tau )$ corresponds to the unique special relativistic
center-of-mass-like worldline of Refs.\cite{beig} \footnote{See
Refs.\cite{israel,ehlers} for the definition of this concept in
general relativity. By using the interpretation of
Ref.\cite{israel}, also the special relativistic limit of the
general relativistic Dixon centroid of Ref.\cite{dixon1} gives the
centroid $x^{\mu}_s(\tau )$: it coincides with  the special
relativistic Dixon centroid of Ref.\cite{dixon} defined by using
the conserved energy momentum tensor, as we shall see  in Section
IV.}, which unifies previous proposals of Synge, M$\o$ller and
Pryce quoted in that paper.

The only remaining canonical variables describing the Wigner
hyperplane in the final Dirac brackets are the non-covariant canonical
coordinate ${\tilde x}^{\mu}_s(\tau )$ \footnote{It describes a point
living on the Wigner hyperplanes and has the covariance of the little
group O(3) of timelike Poincar\'e orbits, like the Newton-Wigner
position operator.} and $p^{\mu}_s$. The point with coordinates
${\tilde x}^{\mu}_s(\tau )$ is the decoupled canonical {\it external
4-center of mass} of the isolated system, which can be interpreted as
a decoupled observer with his parametrized clock ({\it point particle
clock}). Its velocity ${\dot {\tilde x}}^{\mu}_s(\tau )$ is parallel
to $p^{\mu}_s$, so that it has no classical zitterbewegung.

The relation between $x^{\mu}_s(\tau )$ and ${\tilde
x}_s^{\mu}(\tau )$ (${\vec {\tilde \sigma}}$ is its 3-location on
the Wigner hyperplane) is \cite{lus,iten1}

\begin{equation}
{\tilde x}^{\mu}_s(\tau )= ({\tilde x}^o_s(\tau ); {\vec {\tilde
x}}_s(\tau ) )=z^{\mu}(\tau ,{\tilde {\vec \sigma }})=x^{\mu}_s(\tau
)-{1\over {\epsilon_s(p^o_s+\epsilon_s)}}\Big[
p_{s\nu}S_s^{\nu\mu}+\epsilon_s(S^{o\mu}_s-S^{o\nu}_s{{p_{s\nu}p_s^{\mu}}\over
{\epsilon^2_s}}) \Big],
\label{II8}
\end{equation}

After the separation of the relativistic canonical non-covariant {\it
external} 4-center of mass ${\tilde x}_s^{\mu}(\tau )$, on the Wigner
hyperplane the N particles are described by the 6N Wigner spin-1
3-vectors ${\vec \eta}_i(\tau )$, ${\vec \kappa}_i (\tau )$ restricted
by the rest-frame condition ${\vec \kappa}_{+}=\sum^N
_{i=1} {\vec \kappa}_i \approx 0$.

Inside the  Wigner hyperplane, three  degrees of freedom of the
isolated system \footnote{As already said, they describe an {\it
internal} center-of-mass 3-variable ${\vec \sigma}_{com}$ defined
inside the Wigner hyperplane and conjugate to ${\vec \kappa}_{+}$;
when the ${\vec \sigma}_{com}$ are canonical variables they are
denoted ${\vec q}_{+}$.} become gauge variables. To eliminate the
three first class constraints ${\vec \kappa}_{+} \approx 0$ the
natural gauge fixing  is $\vec \chi (\tau )={\vec \sigma}_{com}
={\vec q}_{+}\approx 0$ such that ${\vec q}_{+}\approx 0$  implies
$\lambda_{\check r}(\tau )=0$: in this way the {\it internal}
3-center of mass is located at the origin $x^{\mu}_s(\tau
)=z^{\mu}(\tau ,\vec \sigma =0)$ of the Wigner hyperplane.

The various spin tensors and vectors are \cite{lus}

\begin{eqnarray}
J^{\mu\nu}_s&=&x^{\mu}_s p^{\nu}_s- x^{\nu}_s p^{\mu}_s+ S^{\mu\nu}_s=
{\tilde x}^{\mu}_s p^{\nu}_s - {\tilde x}^{\nu}_s p^{\mu}_s +{\tilde S}
^{\mu\nu}_s,\nonumber \\
&&{}\nonumber \\
S^{\mu\nu}_s&=&[u^{\mu}(p_s)\epsilon^{\nu}(u(p_s))-u^{\nu}(p_s)\epsilon^{\mu}
(u(p_s))] {\bar S}^{\tau r}_s+\epsilon^{\mu}(u(p_s))\epsilon^{\nu}(u(p_s))
{\bar S}^{rs}_s\equiv \nonumber \\
&\equiv& \Big[ \epsilon^{\mu}_r(u(p_s)) u^{\nu}(p_s)-\epsilon^{\nu}
(u(p_s)) u^{\mu}(p_s)\Big] \sum_{i=1}^N \eta^r_i \sqrt{m^2_ic^2+{\vec \kappa}
_i^2}+\nonumber \\
&+&\Big[ \epsilon^{\mu}_r(u(p_s)) \epsilon^{\nu}_s(u(p_s))-\epsilon^{\nu}_r
(u(p_s)) \epsilon^{\mu}_r(u(p_s))\Big] \sum_{i=1}^N \eta^r_i\kappa^s_i,
\nonumber \\
&&{}\nonumber \\
{\bar S}^{AB}_s&=&\epsilon^A_{\mu}(u(p_s)) \epsilon^B_{\nu}(u(p_s)) S^{\mu\nu}_s
,\nonumber \\
&&{\bar S}^{rs}_s\equiv \sum_{i=1}^N(\eta^r_i\kappa^s_i-
\eta^s_i\kappa^r_i),\quad {\bar S}^{\tau r}_s\equiv -
\sum_{i=1}^N\eta^r_i\sqrt{m^2_ic^2+{\vec \kappa}_i^2},\nonumber \\
&&{}\nonumber \\ {\tilde S}^{\mu\nu}_s&=&S^{\mu\nu}_s+{1\over
{\sqrt{\epsilon p^2_s}(p^o_s+
\sqrt{\epsilon p^2_s})}}\Big[ p_{s\beta}(S^{\beta\mu}_s p^{\nu}_s-S^{\beta\nu}_s
p^{\mu}_s)+\sqrt{p^2_s}(S^{o\mu}_s p^{\nu}_s-S^{o\nu}_s p^{\mu}_s)\Big],
\nonumber \\
&&{\tilde S}^{ij}_s=\delta^{ir}\delta^{js} {\bar S}_s^{rs},\quad\quad
{\tilde S}^{oi}_s=-{{\delta^{ir} {\bar S}^{rs}_s\, p^s_s}\over {p^o_s+
\sqrt{\epsilon p^2_s}}},\nonumber \\
&&{}\nonumber \\ {\vec  {\bar S}} &\equiv & {\vec {\bar
S}}=\sum_{i+1}^N{\vec
\eta}_i\times {\vec
\kappa}_i\approx
\sum_{i=1}^N {\vec \eta}_i\times {\vec \kappa}_i-{\vec \eta}_{+}\times
{\vec \kappa}_{+} = \sum_{a=1}^{N-1} {\vec \rho}_a\times {\vec \pi}_a.
\label{II9}
\end{eqnarray}

Let us remark that while
$L^{\mu\nu}_s=x^{\mu}_sp^{\nu}_s-x^{\nu}_sp^{\mu}_s$ and
$S^{\mu\nu}_s$ are not constants of the motion due to classical
zitterbewung, both ${\tilde L}_s^{\mu\nu}={\tilde
x}^{\mu}_sp_s^{\nu}-{\tilde x}^{\nu}_sp^{\mu}_s$ and ${\tilde
S}^{\mu\nu}_s$ are conserved.

The canonical variables ${\tilde x}^{\mu}_s$, $p^{\mu}_s$ for the
{\it external} 4-center of mass, can be replaced by the canonical
pairs\cite{ll}\footnote{It makes explicit the interpretation as a
{\it point particle clock}.}

\begin{eqnarray}
T_s&=& {{p_s\cdot {\tilde x}_s}\over {\epsilon_s}}={{p_s
 \cdot x_s}\over {\epsilon_s}},\nonumber \\
 \epsilon_s&=&\pm \sqrt{\epsilon p^2_s},\nonumber \\
 {\vec z}_s&=&\epsilon_s ({\vec {\tilde
x}}_s- {{ {\vec p}_s}\over {p^o_s}} {\tilde x}^o_s), \nonumber \\
 {\vec k}_s&=&{{ {\vec p}_s}\over {\epsilon_s}},
\label{II10}
\end{eqnarray}

\noindent with the inverse transformation

\begin{eqnarray}
{\tilde x}^o_s&=&\sqrt{1+{\vec k}
_s^2}(T_s+{{{\vec k}_s\cdot {\vec z}_s}\over {\epsilon_s}}), \nonumber \\
 {\vec {\tilde x}}_s&=&{{{\vec z}_s}\over {\epsilon_s}}+(T_s+{{{\vec
k}_s\cdot {\vec z}_s}\over {\epsilon_s}}){\vec k}_s, \nonumber \\
 p^o_s&=&\epsilon_s \sqrt{1+{\vec k}_s^2}, \nonumber \\
 {\vec p}_s&=&\epsilon_s {\vec k}_s.
\label{II11}
\end{eqnarray}

This non-point canonical transformation can be summarized as
[$\epsilon_s - M_{sys} \approx 0$, ${\vec \kappa}_{+}=\sum_{i=1}^N
{\vec \kappa}_i \approx 0$]

\begin{equation}
\begin{minipage}[t]{5cm}
\begin{tabular}{|l|l|} \hline
${\tilde x}_s^{\mu}$& ${\vec \eta}_i$ \\  \hline
 $p^{\mu}_s$& ${\vec \kappa}_i$ \\ \hline
\end{tabular}
\end{minipage} \ {\longrightarrow \hspace{1cm}} \
\begin{minipage}[t]{5 cm}
\begin{tabular}{|l|l|l|} \hline
$\epsilon_s$& ${\vec z}_s$   & ${\vec \eta}_i$   \\ \hline
 $T_s$ & ${\vec k}_s$& ${\vec \kappa}_i$ \\ \hline
\end{tabular}
\end{minipage}
\label{II12}
\end{equation}

The invariant mass $M_{sys}$ of the system, which is also the {\it
internal} energy of the isolated system, replaces the
non-relativistic Hamiltonian $H_{rel}$ for the relative degrees of
freedom, after the addition of the gauge-fixing $T_s-\tau \approx
0$ \footnote{It implies $\lambda (\tau )=-1$ and identifies the
time parameter $\tau$ with the Lorentz scalar time of the center
of mass in the rest frame, $T_s=p_s\cdot {\tilde x}_s/M_{sys}$;
$M_{sys}$  generates the evolution in this time.}: this reminds of
the frozen Hamilton-Jacobi theory, in which the time evolution can
be reintroduced by using the energy generator of the Poincar\'e
group as Hamiltonian \footnote{See Refs.\cite{pon} for a different
derivation of this result.}.

After the gauge fixings $T_s-\tau \approx 0$, the final Hamiltonian
and the embedding of the Wigner hyperplane into Minkowski spacetime
are

\bea
 H_D&=& M_{sys} -\vec \lambda (\tau ) \cdot {\vec \kappa}_{+} ,\nonumber \\
 &&{}\nonumber \\
z^{\mu}(\tau ,\vec \sigma ) &=& x^{\mu}_s(\tau ) + \epsilon^{\mu
}_r(u(p_s)) \sigma^r = x^{\mu}_s(0) + u^{\mu}(p_s) \tau +
\epsilon^{\mu}_r(u(p_s)) \sigma^r,\nonumber \\
 &&{}\nonumber \\
 &&with \nonumber \\
&&{}\nonumber \\
 {\dot x}^{\mu}_s(\tau )\, &{\buildrel \circ \over =}& {{d\, x^{\mu}_s(\tau )}\over
 {d \tau}}+ \{ x^{\mu}_s(\tau ), H_D \} = u^{\mu}(p_s)+\epsilon^{\mu}_r(u(p_s))
\lambda_r(\tau ),
\label{II13}
\eea

\noindent where $x^{\mu}_s(0)$ is an arbitrary point.

The particles' worldlines in Minkowski spacetime and the associated
momenta are

\begin{eqnarray}
x^{\mu}_i(\tau )&=&z^{\mu}(\tau ,{\vec \eta}_i(\tau ))=x^{\mu}_s(\tau
)+\epsilon^{\mu}_r(u(p_s)) \eta^r_i(\tau ),\nonumber \\
 p^{\mu}_i(\tau )&=&\sqrt{m^2_i+{\vec \kappa}^2_i(\tau )} u^{\mu}(p_s) +
\epsilon^{\mu}_r(u(p_s)) \kappa_{ir}(\tau )\,\, \Rightarrow \epsilon p^2_i=m_i^2.
\label{II14}
\end{eqnarray}

The {\it external} rest-frame instant form realization of the
Poincar\'e generators \footnote{As in every instant form of
dynamics, there are four independent Hamiltonians $p^o_s$ and
$J^{oi}_s$, functions of the invariant mass $M_{sys}$; we give
also the expression in the basis $T_s$, $\epsilon_s$, ${\vec
z}_s$, ${\vec k}_s$.} with non-fixed invariants $\epsilon p^2_s =
\epsilon_s^2 \approx M^2_{sys}$, $-\epsilon p^2_s {\vec {\bar
S}}_s^2 \approx -\epsilon M^2_{sys} {\vec {\bar S}}^2$, is
obtained from Eq.(\ref{II9}):

\begin{eqnarray}
p^{\mu}_s,&&\nonumber  \\
 J^{\mu\nu}_s&=&{\tilde x}^{\mu}_sp^{\nu}_s-{\tilde x}^{\nu}_s p^{\mu}_s
 + {\tilde S}^{\mu\nu}_s,\nonumber \\
 &&{}\nonumber \\
 p^o_s&=& \sqrt{\epsilon_s^2+{\vec p}_s^2}= \epsilon_s \sqrt{1+ {\vec
k}_s^2}\approx  \sqrt{M^2_{sys}+{\vec p}^2_s}=M_{sys} \sqrt{1+{\vec
k}_s^2},\nonumber \\
 {\vec p}_s&=& \epsilon_s{\vec k}_s\approx M_{sys} {\vec k}_s,
\nonumber \\
 J^{ij}_s&=&{\tilde x}^i_sp^j_s-{\tilde x}^j_sp^i_s +
\delta^{ir}\delta^{js}\sum_{i=1}^N(\eta^r_i\kappa^s_i-\eta^s_i\kappa^r_i)=
z^i_sk^j_s-z^j_sk^i_s+\delta^{ir}\delta^{js} \epsilon^{rsu}{\bar
S}^u_s,\nonumber \\
 K^i_s&=&J^{oi}_s= {\tilde x}^o_sp^i_s-{\tilde x}^i_s
\sqrt{\epsilon_s^2+{\vec p}_s^2}-{1\over {\epsilon_s+\sqrt{\epsilon^2_s+
{\vec p}_s^2}}} \delta^{ir} p^s_s \sum_{i=1}^N(\eta^r_i\kappa^s_i-
\eta^s_i\kappa^r_i)=\nonumber \\
 &=&-\sqrt{1+{\vec k}_s^2} z^i_s-{{\delta^{ir} k^s_s\epsilon
^{rsu}{\bar S}^u_s}\over {1+\sqrt{1+{\vec k}_s^2} }}\approx {\tilde x}^o_sp^i_s
-{\tilde x}^i_s\sqrt{M^2_{sys}+{\vec p}^2_s}-{{\delta^{ir}p^s_s\epsilon^{rsu}{\bar
S}^u_s}\over {M_{sys}+\sqrt{M_{sys}^2+{\vec p}^2_s}}}.
\label{II15}
\end{eqnarray}

On the other hand the {\it internal} realization of the Poincar\'e
algebra is built inside the Wigner hyperplane by using the
expression of ${\bar S}_s^{AB}$ given by Eq.(\ref{II9})
\footnote{This {\it internal Poincar\'e algebra realization}
should not be confused with the previous {\it external} one based
on ${\tilde S}^{\mu\nu}_s$; $\Pi$ and $W^2$ are the two non-fixed
invariants of this realization.}

\begin{eqnarray}
&&M_{sys}=H_M=\sum_{i=1}^N  \sqrt{m^2_i+{\vec
\kappa}_i^2},\nonumber \\
 &&{\vec \kappa}_{+}=\sum_{i=1}^N {\vec
\kappa}_i\, (\approx 0),\nonumber \\
 &&\vec J=\sum_{i=1}^N {\vec \eta}_i\times {\vec
\kappa}_i,\quad\quad J^r={\bar S}^r={1\over 2}\epsilon^{ruv}{\bar
S}^{uv} \equiv {\bar S}^r_s,\nonumber \\
 &&\vec K=- \sum_{i=1}^N
\sqrt{m^2_i+{\vec \kappa}_i^2}\,\, {\vec
\eta}_i=-M_{sys}{\vec R}_{+},
 \quad\quad K^r=J^{or}={\bar S}_s^{\tau r}, \nonumber \\
&&{}\nonumber \\
 &&\Pi = M^2_{sys}-{\vec \kappa}_{+}^2 \approx M^2_{sys} > 0,\nonumber \\
 &&W^2=-\epsilon (M^2_{sys}-{\vec \kappa}^2_{+}) {\vec {\bar
S}}^2_s \approx -\epsilon M^2_{sys} {\vec {\bar S}}^2_s.
\label{II16}
\end{eqnarray}

The constraints $\epsilon_s-M_{sys}\approx 0$, ${\vec
\kappa}_{+}\approx 0$ mean: i) the constraint
$\epsilon_s-M_{sys}\approx 0$ is the bridge which connects the
{\it external} and {\it internal} realizations \footnote{The
external spin coincides with the internal angular momentum due to
Eqs.(\ref{a11}) of Ref.\cite{iten1}.}; ii) the constraints ${\vec
\kappa}_{+}\approx 0$, together with $\vec K\approx 0$
\footnote{As we shall see in the next Section, $\vec K\approx 0$
is implied by the natural gauge fixing ${\vec q}_{+}\approx 0$.},
imply a unfaithful {\it internal} realization, in which the only
non-zero generators are the conserved energy and spin of an
isolated system.

The determination of ${\vec q}_{+}$  for the N particle system was
done with the group theoretical methods of Ref.\cite{pauri} in
Section III of Ref.\cite{iten1}. Given a realization  of the ten
Poincar\'e generators on the phase space, one can build three
3-position variables in terms of them only. For N free scalar
relativistic particles on the Wigner hyperplane with ${\vec
p}_{sys}={\vec \kappa}_{+}\approx 0$ and by using the {\it
internal} realization (\ref{II16}) they are:\hfill\break
 i) a canonical {\it internal} center of mass (or center of spin)  ${\vec q}_{+}$;
\hfill\break
ii) a non-canonical {\it internal}  M\o ller center of energy ${\vec
R}_{+}$;\hfill\break
 iii) a non-canonical {\it internal} Fokker-Pryce
center of inertia ${\vec y}_{+}$. \hfill\break
 It can be shown\cite{iten1} that, due to ${\vec \kappa}_{+}\approx 0$, they
 {\it coincide}: ${\vec q}_{+} \approx {\vec R}_{+} \approx {\vec y}_{+}$.

Therefore the gauge fixings $\vec \chi (\tau )={\vec q}_{+}\approx
{\vec R}_{+}\approx {\vec y}
_{+}\approx 0$ imply $\vec \lambda (\tau )\approx 0$ and force the three
{\it internal} collective variables to coincide with the origin of the
coordinates, which now becomes

\beq
x^{({\vec q}_{+})\mu}_s(T_s)=x_s^{\mu}(0) +u^{\mu}(p_s) T_s.
\label{II17}
\eeq

\noindent As  shown in Section IV, the addition of the gauge
fixings $\vec \chi (\tau )={\vec q}_{+}\approx {\vec R}_{+}\approx
{\vec y}_{+} \approx 0$  implies that the Dixon center of mass of
an extended object\cite{dixon1} and  the Pirani\cite{pirani} and
Tulczyjew\cite{mul4,ehlers,mul8} centroids \footnote{See Ref.
\cite{mate} for the application of these methods to find the
center of mass of a configuration of the Klein-Gordon field after
the preliminary work of Ref.\cite{lon} on the {\it center of
phase} for a real Klein-Gordon field.} all simultaneously coincide
with  the origin $x_s^{\mu }(\tau )$.

The {\it external} realization (\ref{II15}) allows to build the
analogous {\it external} 3-variables ${\vec q}_s$, ${\vec R}_s$,
${\vec Y}_s$. It is then shown in Ref.\cite{iten1} how to build
the associated {\it external} 4-variables and their location on
the Wigner hyperplane

\begin{eqnarray}
{\tilde x}^{\mu}_s&=&( {\tilde x}^o_s; {\vec {\tilde x}}_s)= ({\tilde
x}^o_s; {\vec q}_s+{{{\vec p}_s}\over {p^o_s}} {\tilde
x}^o_s)=\nonumber \\
 &=&({\tilde x}^o_s; {{{\vec z}_s}\over {\epsilon_s}}+(T_s+{{{\vec
k}_s\cdot {\vec z}_s}\over {\epsilon_s}}){\vec k}_s )
=x^{\mu}_s+\epsilon^{\mu}_u(u(p_s)) {\tilde \sigma}^u,
\nonumber \\
 Y^{\mu}_s&=&({\tilde x}^o_s; {\vec Y}_s)=\nonumber \\
  &=&({\tilde x}^o_s;\, {1\over {\epsilon_s}}[{\vec z}_s+{{{\vec
{\bar S}}_s\times {\vec p}_s}\over {\epsilon_s[1+u^o(p_s)]}}]+(T_s+
{{{\vec k}_s\cdot {\vec z}_s}\over {\epsilon_s}}){\vec k}_s\,
)=\nonumber \\
 &=&{\tilde x}^{\mu}_s+\eta^{\mu}_r{{({\vec {\bar
S}}_s\times {\vec p}_s)^r}\over {\epsilon_s[1+u^o(p_s)]}}=\nonumber \\
 &=&x^{\mu}_s+\epsilon^{\mu}_u(u(p_s)) \sigma^u_Y,\nonumber \\
 R^{\mu}_s&=&({\tilde x}^o_s; {\vec R}_s)=\nonumber \\
  &=&( {\tilde x}^o_s;\, {1\over {\epsilon_s}}[{\vec z}_s-
{{{\vec {\bar S}}_s\times {\vec p}_s}\over {\epsilon_s u^o(p_s)
[1+u^o(p_s)]}}]+(T_s+ {{{\vec k}_s\cdot {\vec z}_s}\over
{\epsilon_s}}){\vec k}_s\, )=\nonumber \\
 &=&{\tilde
x}^{\mu}_s-\eta^{\mu}_r{{({\vec {\bar S}}_s\times {\vec p}_s)^r}
\over {\epsilon_su^o(p_s)[1+u^o(p_s)]}}=\nonumber \\
&=&x^{\mu}_s+\epsilon^{\mu}_u(u(p_s)) \sigma^u_R,\nonumber \\
&&{}\nonumber \\
T_s&=&u(p_s)\cdot x_s=u(p_s)\cdot {\tilde x}_s=u(p_s)\cdot Y_s=u(p_s)\cdot
R_s,\nonumber \\
&&{}\nonumber \\
{\tilde \sigma}^r&=&\epsilon_{r\mu}(u(p_s))[x^{\mu}_s-{\tilde x}^{\mu}_s]=
{{ \epsilon_{r\mu}(u(p_s)) [u_{\nu}(p_s)S^{\nu\mu}_s+S^{o\mu}_s]}\over
{[1+u^o(p_s)]}}=\nonumber \\
&=&-{\bar S}_s^{\tau r}+{{{\bar S}_s^{rs}p^s_s}\over {\epsilon_s[1+u^o(p_s)]}}
=\epsilon_s R^r_{+}+{{{\bar S}_s^{rs}u^s(p_s)}\over {1+u^o(p_s)}}
\approx \nonumber \\
&\approx& \epsilon_s q^r_{+}+{{{\bar S}_s^{rs}u^s(p_s)}\over {1+
u^o(p_s)}}\approx {{{\bar S}_s^{rs}u^s(p_s)}\over {1+u^o(p_s)}}
,\nonumber \\
\sigma^r_Y&=&\epsilon_{r\mu}(u(p_s))[x^{\mu}_s-Y^{\mu}_s]={\tilde \sigma}^r-
\epsilon_{ru}(u(p_s)){{({\vec
{\bar S}}_s\times {\vec p}_s)^u}\over {\epsilon_s[1+u^o(p_s)]}}=\nonumber \\
&=&{\tilde \sigma}^r+{{{\bar S}^{rs}_su^s(p_s)}\over {1+u^o(p_s)}}=
\epsilon_s R^r_{+} \approx \epsilon_s q^r_{+} \approx 0,\nonumber \\
\sigma^r_R&=&\epsilon_{r\mu}(u(p_s))[x^{\mu}_s-R^{\mu}_s]={\tilde \sigma}^r+
\epsilon_{ru}(u(p_s)) {{({\vec {\bar S}}_s\times {\vec p}_s)^u}
\over {\epsilon_su^o(p_s)[1+u^o(p_s)]}}=\nonumber \\
&=&{\tilde \sigma}^r-{{{\bar S}_s^{rs}u^s(p_s)}\over {u^o(p_s)[1+
u^o(p_s)]}}=\epsilon_sR^r_{+}+{{[1-u^o(p_s)]{\bar S}^{rs}_su^s(p_s)}\over
{u^o(p_s)[1+u^o(p_s)]}}\approx \nonumber \\
&\approx& {{[1-u^o(p_s)]{\bar S}^{rs}_su^s(p_s)}\over
{u^o(p_s)[1+u^o(p_s)]}},\nonumber \\
&&{}\nonumber \\
&\Rightarrow& x^{({\vec q}_{+})\mu}_s(\tau ) = Y^{\mu}_s,
\label{II18}
\end{eqnarray}

\noindent namely the {\it external} Fokker-Pryce non-canonical center of inertia
coincides with the origin $x^{({\vec q}_{+})\mu}_s(\tau )$ carrying
the internal center of mass.

\vfill\eject

\section{The Energy-Momentum Tensor on the Wigner Hyperplane and Dixon's
Relativistic Multipoles.}

\subsection{The Euler-Lagrange Equations and the Energy-Momentum Tensor
of Parametrized Minkowski Theories.}

The Euler-Lagrange equations associated with the Lagrangian
(\ref{II1}) are  (the symbol '${\buildrel \circ \over =}$' means
evaluated on the solutions of the equations of motion)

\begin{eqnarray}
&&\Big( {{\partial {\cal L}}\over {\partial z^{\mu}}}-\partial_A
{{\partial {\cal L}}\over {\partial z^{\mu}_A}}\Big) (\tau ,\vec
\sigma )=\eta_{\mu\nu}\partial_A[\sqrt{g} T^{AB} z_B^{\nu}](\tau ,\vec
\sigma )\, {\buildrel \circ \over =}\, 0,\nonumber \\
 &&{{\partial L}\over
{\partial {\vec \eta}_i}}-\partial_{\tau} {{\partial L}\over {\partial
{\dot {\vec \eta}}_i}}=\nonumber \\
 &&-[{1\over 2} {{T^{AB}}\over {\sqrt{g}}}]{|}_{\vec \sigma ={\vec \eta}_i}
 {{\partial g_{AB}}\over {\partial{\vec \eta}_i}} -\partial_{\tau}
 {{ g_{\tau r}+g_{rs}{\dot \eta}_i^s}\over {\sqrt{g_{\tau\tau}+2g_{\tau u}{\dot \eta}^u_i
 +g_{uv}{\dot \eta}^u_i{\dot \eta}^v_i}}}{|}_{\vec \sigma ={\vec \eta}_i}
\, {\buildrel \circ \over =}\, 0,
\label{III1}
\end{eqnarray}

\noindent where we have introduced the energy-momentum tensor
[here ${\dot \eta}^A_i(\tau )=(1; {\dot {\vec \eta}}_i(\tau ))$]

\begin{equation}
T^{AB}(\tau ,\vec \sigma )=-[{2\over {\sqrt{g}}}{{\delta S}\over
{\delta g_{AB}}}](\tau ,\vec \sigma )=-\sum_{i=1}^N\delta^3(\vec
\sigma -{\vec \eta}_i(\tau )) {{m_i{\dot \eta}^A_i(\tau ){\dot \eta}^B_i(\tau )}\over
{\sqrt{g_{\tau\tau}+2g_{\tau u}{\dot \eta}^u_i
 +g_{uv}{\dot \eta}^u_i{\dot \eta}^v_i}}}(\tau ,\vec \sigma ).
\label{III2}
\end{equation}

Due to the delta functions the Euler-Lagrange equations for the
fields $z^{\mu}(\tau ,\vec \sigma )$ are trivial ($0\, {\buildrel
\circ \over =}\, 0$) everywhere except at the positions of the
particles. They may be rewritten in a form valid for every
isolated system

\begin{equation}
 \partial_AT^{AB} z^{\mu}_B\, {\buildrel \circ \over =}\, -{1\over {\sqrt{g}}}
 \partial_A[\sqrt{g} z^{\mu}_B] T^{AB}.
\label{III3}
\end{equation}

\noindent When $\partial_A[\sqrt{g} z^{\mu}_B]=0$ as it happens on
the Wigner hyperplanes in the gauge ${\vec q}_{+}\approx 0$ and
$T_s-\tau \approx 0$, we get the conservation of the
energy-momentum tensor $T^{AB}$, i.e. $\partial_AT^{AB}\,
{\buildrel \circ \over =}\, 0$. Otherwise there is a compensation
coming from the dynamics of the surface.

On the Wigner hyperplane, where we have

\begin{eqnarray}
x^{\mu}_i(\tau )&=&z^{\mu}(\tau ,{\vec \eta}_i(\tau ))=x^{\mu}_s(\tau
)+\epsilon^{\mu}_r(u(p_s))\eta^r_i(\tau ),\nonumber \\
 &&{\dot x}^{\mu}_i(\tau )=z^{\mu}_{\tau}(\tau ,{\vec \eta}_i(\tau
))+z^{\mu}_r(\tau ,{\vec \eta}_i(\tau)) {\dot \eta}^r_i(\tau )={\dot
x}_s^{\mu}(\tau )+\epsilon^{\mu}_r(u(p_s)) {\dot \eta}^r_i(\tau
),\nonumber \\
 &&{\dot x}^2_i(\tau )=g_{\tau\tau}(\tau ,{\vec
\eta}_i(\tau ))+2g_{\tau r}(\tau ,{\vec \eta}_i(\tau )){\dot \eta}^r
_i(\tau )+g_{rs}(\tau ,{\vec \eta}_i(\tau )){\dot \eta}_i^r(\tau ){\dot \eta}
^s_i(\tau )=\nonumber \\
 &&{\dot x}_s^2(\tau )+2{\dot x}_{s\mu}(\tau
)\epsilon^{\mu}_r(u(p_s)){\dot \eta}^r_i(\tau )- {\dot {\vec
\eta}}_i^2(\tau ),\nonumber \\
 &&{}\nonumber \\
p^{\mu}_i(\tau )&=&\sqrt{m^2_i-\gamma^{rs}(\tau ,{\vec \eta}
_i(\tau ))\kappa_{ir}(\tau )\kappa_{is}(\tau )} l^{\mu}(\tau ,{\vec \eta}
_i(\tau ))-\kappa_{ir}(\tau )\gamma^{rs}(\tau ,{\vec \eta}_i(\tau )) z^{\mu}_s
(\tau ,{\vec \eta}_i(\tau ))=\nonumber \\
 &=&\sqrt{m_i^2+{\vec \kappa}_i^2(\tau )}u^{\mu}(p_s)+
 \epsilon^{\mu}_r(u(p_s))\kappa_i^r(\tau )\,\, \Rightarrow p^2_i=m^2_i, \nonumber \\
 p^{\mu}_s&=&\int d^3\sigma \rho_{\mu}(\tau ,\vec \sigma )\approx
\sum_{i=1}^N p^{\mu}_i(\tau ),
\label{III4}
\end{eqnarray}

\noindent the energy-momentum tensor $T^{AB}(\tau ,\vec \sigma )$ has the form

\begin{eqnarray}
T^{\tau\tau}(\tau ,\vec \sigma )&=&-\sum_{i=1}^N \delta^3(\vec \sigma
-{\vec \eta}_i(\tau )) {{m_i}\over {\sqrt{{\dot x}_s^2(\tau )+
2{\dot x}_{s\mu}(\tau )\epsilon^{\mu}_r(u(p_s))-{\dot {\vec
\eta}}_i^2(\tau )}}},\nonumber \\
T^{\tau r}(\tau ,\vec \sigma )&=&-\sum_{i=1}^N \delta^3(\vec \sigma
-{\vec \eta}_i(\tau )) {{m_i{\dot \eta}^r_i(\tau )}\over {\sqrt{{\dot x}_s^2(\tau )+
2{\dot x}_{s\mu}(\tau )\epsilon^{\mu}_r(u(p_s))-{\dot {\vec
\eta}}_i^2(\tau )}}},\nonumber \\
T^{rs}(\tau ,\vec \sigma )&=&-\sum_{i=1}^N \delta^3(\vec \sigma
-{\vec \eta}_i(\tau )) {{m_i{\dot \eta}_i^r(\tau )
{\dot \eta}_i^s(\tau )}\over {\sqrt{{\dot x}_s^2(\tau )+ 2{\dot
x}_{s\mu}(\tau )\epsilon^{\mu}_r(u(p_s))-{\dot {\vec
\eta}}_i^2(\tau )}}}.
\label{III5}
\end{eqnarray}

\subsection{The Energy-Momentum Tensor of the Standard Lorentz-Covariant Theory.}

The same form is obtained from the restriction to positive energies
\footnote{$p^o_i=m_i {{ {\dot x}^o_i}\over {\sqrt{{\dot x}^2_i} }} > 0$,
$\epsilon (\sum_{i=1}^Np_i^{\mu})^2=\epsilon (\sum_{i=1}^Nm_i{{ {\dot
x}_i^{\mu}}\over {\sqrt{{\dot x}^2_i} }})^2 > 0$.} of the energy
momentum tensor of the standard manifestly Lorentz covariant theory
with Lagrangian $S_S=\int d\tau L_S(\tau )=-\sum_{i=1}^Nm_i \int d\tau
\sqrt{{\dot x}^2_i(\tau )}$. On the Wigner hyperplanes with $T_s-\tau
\approx 0$ we will get

\begin{eqnarray}
T^{\mu\nu}(\tau ,\vec \sigma )&=&- \Big( {2\over {\sqrt{g}}}\,
{{\delta S_S}\over {\delta g_{\mu\nu}}}\Big) {|}_{x=z(\tau ,\vec
\sigma )}=\nonumber \\
 &=&\sum_{i=1}^N m_i
\int d\tau_1 {{ {\dot x}_i^{\mu}(\tau_1){\dot x}_i^{\nu}(\tau_1)}\over
{\sqrt{{\dot x}^2_i(\tau_1)}}} \delta^4\Big(x_i(\tau_1)-z(\tau ,\vec
\sigma )\Big)=\nonumber \\
 &=&\epsilon^{\mu}_A(u(p_s))\epsilon^{\nu}_B(u(p_s)) T^{AB}(\tau ,\vec
\sigma ).
\label{III6}
\end{eqnarray}

On the other hand, from the restriction of the standard theory we
get \hfill\break \hfill\break

1) On arbitrary {\it spacelike hypersurfaces}

\begin{eqnarray}
T^{\mu\nu}(z(\tau ,\vec \sigma ))&=&\sum_{i=1}^N m_i
\int d\tau_1 {{ {\dot x}_i^{\mu}(\tau_1){\dot x}_i^{\nu}(\tau_1)}\over
{\sqrt{{\dot x}^2_i(\tau_1)}}} \delta^4\Big(x_i(\tau_1)-z(\tau ,\vec
\sigma )\Big)=\nonumber \\
&=&\sum_{i=1}^N m_i
 \int {{d\tau_1 }\over {\sqrt{{\dot x}^2_i(\tau_1)}}} \delta^4\Big(z(\tau_1,
{\vec \eta}_i(\tau_1) )-z(\tau ,\vec \sigma )\Big)\nonumber \\
&&[z^{\mu}_{\tau}(\tau_1,{\vec \eta}
_i(\tau_1))+z^{\mu}_r(\tau_1,{\vec \eta}_i(\tau_1)){\dot \eta}
^r_i(\tau_1)]\nonumber \\
&& [z^{\nu}_{\tau}(\tau_1,{\vec \eta}_i(\tau_1))+z^{\nu}_r(\tau_1
,{\vec \eta}_i(\tau_1)){\dot \eta}^r_i(\tau_1)] =\nonumber \\
&=&\sum_{i=1}^N m_i
\int {{d\tau_1}\over {\sqrt{{\dot x}^2_i(\tau_1)}}}
\delta^4\Big( z(\tau_1,{\vec \eta}_i(\tau_1) )-z(\tau ,\vec \sigma )\Big)
\nonumber \\
&& \Big[ z^{\mu}_{\tau}(\tau_1,{\vec \eta}
_i(\tau_1))\, z^{\nu}_{\tau}(\tau_1,{\vec \eta}_i(\tau_1))+\nonumber \\
&+&\Big(
z^{\mu}_{\tau}(\tau_1,{\vec \eta}_i(\tau_1))z^{\nu}_r(\tau_1,{\vec \eta}
_i(\tau_1))+z^{\nu}_{\tau}(\tau_1,{\vec \eta}_i(\tau_1))z^{\mu}_r(\tau_1,{\vec
\eta}_i(\tau_1))\Big) {\dot \eta}_i^r(\tau_1)+\nonumber \\
&+&z^{\mu}_r(\tau_1,{\vec \eta}_i(\tau_1))
z^{\nu}_s(\tau_1,{\vec \eta}_i(\tau_1))
{\dot \eta}^r_i(\tau_1) {\dot \eta}^s_i(\tau_1) \Big]
=\nonumber \\
&=&\sum_{i=1}^N {{m_i}\over {\sqrt{{\dot x}^2_i(\tau )}}}
\delta^3\Big( \vec \sigma -{\vec \eta}_i(\tau )\Big)
(det\, |z^{\mu}_A(\tau ,\vec \sigma )|)^{-1}\nonumber \\
&&\Big[ z^{\mu}_{\tau}(\tau,{\vec \eta}_i(\tau_1))\, z^{\nu}_{\tau}(\tau ,{\vec
\eta}_i(\tau ))+\Big( z^{\mu}_{\tau}(\tau,{\vec \eta}_i(\tau ))
z^{\nu}_r(\tau ,{\vec \eta}_i(\tau ))+\nonumber \\
&+&z^{\nu}_{\tau}(\tau ,{\vec \eta}_i(\tau ))z^{\mu}_r(\tau ,{\vec
\eta}_i(\tau ))\Big) {\dot \eta}_i^r(\tau )+
z^{\mu}_r(\tau ,{\vec \eta}_i(\tau ))
z^{\nu}_s(\tau ,{\vec \eta}_i(\tau ))
{\dot \eta}^r_i(\tau ) {\dot \eta}^s_i(\tau ) \Big] =\nonumber \\
&=&\sum_{i=1}^N {{m_i}\over {\sqrt{{\dot x}^2_i(\tau )}\sqrt{g(\tau ,\vec
\sigma )}}} \delta^3\Big( \vec \sigma -{\vec \eta}_i(\tau )\Big)\nonumber \\
&&\Big[ z^{\mu}_{\tau}(\tau,{\vec \eta}_i(\tau_1))\, z^{\nu}_{\tau}(\tau ,{\vec
\eta}_i(\tau ))+\Big( z^{\mu}_{\tau}(\tau,{\vec \eta}_i(\tau ))
z^{\nu}_r(\tau ,{\vec \eta}_i(\tau ))+\nonumber \\
&+&z^{\nu}_{\tau}(\tau ,{\vec \eta}_i(\tau ))z^{\mu}_r(\tau ,{\vec
\eta}_i(\tau ))\Big) {\dot \eta}_i^r(\tau )+
z^{\mu}_r(\tau ,{\vec \eta}_i(\tau ))
z^{\nu}_s(\tau ,{\vec \eta}_i(\tau ))
{\dot \eta}^r_i(\tau ) {\dot \eta}^s_i(\tau ) \Big] ,
\label{III7}
\end{eqnarray}

\noindent since $det\, |z^{\mu}_A| =\sqrt{g}=\sqrt{\gamma (g_{\tau\tau}-\gamma^{rs}g
_{\tau r}g_{\tau s})}$, $\gamma =|det\, g_{rs}|$.\hfill\break
\hfill\break

2) On arbitrary {\it spacelike hyperplanes}, where it
holds\cite{lus}

\begin{eqnarray}
z^{\mu}(\tau ,\vec \sigma )&=&x^{\mu}_s(\tau )+b^{\mu}_u(\tau
)\sigma^u,\quad\quad x^{\mu}_i(\tau )=x^{\mu}_s(\tau )+ b^{\mu}_u(\tau
)\eta^u_i(\tau ),\nonumber \\
 &&{}\nonumber \\
 z^{\mu}_r(\tau ,\vec \sigma
)&=&b^{\mu}_r(\tau ),\quad\quad z^{\mu}_{\tau}(\tau ,\vec \sigma
)={\dot x}_s^{\mu}(\tau )+{\dot b}^{\mu}_u (\tau
)\sigma^u=l^{\mu}/\sqrt{g^{\tau\tau}}-g_{\tau r}z^{\mu}_r,\nonumber \\
 &&{}\nonumber \\
g_{\tau \tau}&=& [{\dot x}_s^{\mu}+ {\dot
b}^{\mu}_r\sigma^r]^2,\quad\quad g_{\tau r}=b_{r\mu}[{\dot x}_s^{\mu}+
{\dot b}^{\mu}_s\sigma^s],
\nonumber \\
g_{rs}&=&-\delta_{rs},\quad\quad \gamma^{rs}=-\delta^{rs},\quad\quad
\gamma =1,\nonumber \\
g&=&g_{\tau\tau}+\sum_rg^2_{\tau r}, \nonumber \\
  g^{\tau\tau}&=&1/[l_{\mu}({\dot x}_s^{\mu}+{\dot b}^{\mu}_u\sigma^u)]^2,
\quad\quad g^{\tau r}
=g^{\tau\tau}g_{\tau r}=b_{r\mu}({\dot x}_s^{\mu}+{\dot b}^{\mu}_u\sigma^u)/
[l_{\mu}({\dot x}_s^{\mu}+{\dot b}^{\mu}_u\sigma^u)]^2,\nonumber \\
 &&{}\nonumber \\
g^{rs}&=&-\delta^{rs} +g^{\tau\tau}g_{\tau r}g_{\tau s}=-\delta^{rs}+
b_{r\mu}({\dot x}_s^{\mu}+{\dot b}^{\mu}_u\sigma^u) b_{r\nu}({\dot
x}_s^{\nu}+{\dot b}^{\nu}_v\sigma^v)
/[l_{\mu}({\dot x}_s^{\mu}+{\dot b}^{\mu}_u\sigma^u)]^2,
\label{III8}
\end{eqnarray}

\noindent we get

\begin{eqnarray}
T^{\mu\nu}[x^{\beta}_s(\tau )+b^{\beta}_r(\tau )\sigma^r]&=&
\sum_{i=1}^N {{m_i \delta^3\Big( \vec \sigma -{\vec \eta}_i(\tau )\Big) }
\over {\sqrt{g(\tau ,\vec \sigma )}
\sqrt{g_{\tau\tau}(\tau ,\vec \sigma )+2g_{\tau r}(\tau ,\vec \sigma ){\dot
\eta}_i(\tau )-{\dot {\vec \eta}}^2_i(\tau )}  }}\nonumber \\
&&\Big[
({\dot x}^{\mu}_s(\tau )+{\dot b}^{\mu}_r(\tau )\eta^r_i(\tau ))
({\dot x}^{\nu}_s(\tau )+{\dot b}^{\nu}_s(\tau )\eta^s_i(\tau ))+\nonumber \\
&+&\Big( ({\dot x}^{\mu}_s(\tau )+{\dot b}^{\mu}_s(\tau )\eta^s_i(\tau ))
b^{\nu}_r(\tau )+({\dot x}^{\nu}_s(\tau )+{\dot b}^{\nu}_s(\tau )\eta^s
_i(\tau )) b^{\mu}_r(\tau )\Big) {\dot \eta}^r_i(\tau )+\nonumber \\
&+&b^{\mu}_r(\tau ) b^{\nu}_s(\tau ) {\dot \eta}^r_i(\tau ){\dot
\eta}^s_i(\tau ) \Big] .
\label{III9}
\end{eqnarray}

\hfill\break

3) On {\it Wigner's hyperplanes}, where it holds\cite{lus}

\begin{eqnarray}
z^{\mu}(\tau ,\vec \sigma )&=&x^{\mu}_s(\tau
)+\epsilon^{\mu}_u(u(p_s))\sigma^u,\quad\quad x^{\mu}_i(\tau
)=x^{\mu}_s(\tau )+\epsilon^{\mu}_u(u(p_s))
\eta^u_i(\tau ),\nonumber \\
 &&{}\nonumber \\
z^{\mu}_r&=&\epsilon^{\mu}_r(u(p_s),\quad\quad
l^{\mu}=u^{\mu}(p_s),\quad\quad z^{\mu}_{\tau}={\dot x}^{\mu}_s(\tau
),\nonumber \\
 g&=&[{\dot x}_s(\tau )\cdot u(p_s)]^2,\quad\quad g_{\tau\tau}={\dot x}^2_s,
 \quad\quad g_{\tau r}={\dot x}_{s\mu} \epsilon^{\mu}_r(u(p_s)),\quad\quad
 g_{rs}=-\delta_{rs},\nonumber \\
g^{\tau\tau}&=&1/[{\dot x}_{s\mu}u^{\mu}(p_s)]^2,\quad\quad g^{\tau
r}={\dot x}_{s\mu}\epsilon^{\mu}
_r(u(p_s))/[{\dot x}_{s\mu}u^{\mu}(p_s)]^2,\nonumber \\
g^{rs}&=&-\delta^{rs}+{\dot x}
_{s\mu}\epsilon^{\mu}_r(u(p_s)) {\dot x}_{s\nu}\epsilon^{\nu}_s(u(p_s))
/[{\dot x}_{s\mu}u^{\mu}(p_s)]^2,\nonumber \\
 &&{}\nonumber \\
ds^2&=& {\dot x}_s^2(\tau )d\tau^2+2{\dot x}_s(\tau )\cdot
\epsilon_r(u(p_s))d\tau d\sigma^r-d{\vec \sigma}^2,
\label{III10}
\end{eqnarray}

\noindent we get

\begin{eqnarray}
T^{\mu\nu}[x^{\beta}_s(\tau )+\epsilon^{\beta}_r(u(p_s))\sigma^r]&=&
{1\over {{{\dot x}_s(\tau )\cdot u(p_s)} }}
\sum_{i=1}^N {{m_i}\over {\sqrt{{\dot x}_s^2(\tau )+2{\dot x}_{s\beta}(\tau )
\epsilon^{\beta}_r(u(p_s)){\dot \eta}^r_i(\tau )-{\dot {\vec \eta}}_i^2(\tau )}
}}\nonumber \\
&&\delta^3\Big( \vec \sigma -{\vec \eta}
_i(\tau )\Big) \Big[ {\dot x}^{\mu}_s(\tau ){\dot x}^{\nu}_s(\tau )+
\nonumber \\
&+& \Big( {\dot x}^{\mu}_s(\tau )\epsilon^{\nu}_r(u(p_s))+{\dot x}^{\nu}_s(\tau
)\epsilon^{\mu}_r(u(p_s))\Big) {\dot \eta}^r_i(\tau )+\nonumber \\
&+&\epsilon^{\mu}_r(u(p_s))
\epsilon^{\nu}_s(u(p_s)) {\dot \eta}^r_i(\tau ){\dot \eta}^s_i(\tau ) \Big] =
\nonumber \\
&=&{1\over {{{\dot x}_s(\tau )\cdot u(p_s)}}}
\sum_{i=1}^N {{m_i}\over {\sqrt{{\dot x}_s^2(\tau )+2{\dot x}_{s\beta}(\tau )
\epsilon^{\beta}_r(u(p_s)){\dot \eta}^r_i(\tau )-{\dot {\vec \eta}}_i^2(\tau )}
}}\nonumber \\
&&\Big[{\dot x}^{\mu}_s(\tau ){\dot x}^{\nu}
_s(\tau )  \delta^3\Big( \vec \sigma -{\vec \eta}
_i(\tau )\Big)+\nonumber \\
&+&\Big( {\dot x}^{\mu}_s(\tau )\epsilon^{\nu}_r(u(p_s))+{\dot x}^{\nu}_s(\tau
)\epsilon^{\mu}_r(u(p_s))\Big)  \delta^3\Big( \vec \sigma -
{\vec \eta}_i(\tau )\Big) {\dot \eta}^r_i(\tau )+\nonumber \\
&+&\epsilon^{\mu}_r(u(p_s))\epsilon^{\nu}_s(u(p_s))
 \delta^3\Big( \vec \sigma -{\vec \eta}
_i(\tau )\Big) {\dot \eta}^r_i(\tau ){\dot \eta}^s_i(\tau ) \Big] .
\label{III11}
\end{eqnarray}

Since the volume element on the Wigner hyperplane is $u^{\mu}(p_s)
d^3\sigma$, we obtain the following total 4-momentum and total
mass of the N free particle system   (Eqs.(\ref{II7}) are used)

\begin{eqnarray}
P^{\mu}_T&=&\int d^3\sigma T^{\mu\nu}[x^{\beta}_s(\tau )+\epsilon^{\beta}
_u(u(p_s))\sigma^u] u_{\nu}(p_s)=\nonumber \\
&=& \sum_{i=1}^N {{m_i}\over {\sqrt{\lambda^2(\tau )- [{\dot {\vec
\eta}}_i(\tau )+\vec \lambda (\tau )]^2}}}\nonumber \\ &&\Big[
-\lambda (\tau )u^{\mu}(p_s)+ [{\dot
\eta}^r_i(\tau )+\lambda^r(\tau )] \epsilon^{\mu}_r(u(p_s))\Big]  \nonumber \\
&{\buildrel \circ \over =}\,& {|}_{\lambda (\tau )=-1}\,\, \sum_{i=1}^N\Big [
\sqrt{m^2_ic^2+{\vec \kappa}_i^2(\tau )}u^{\mu}(p_s)+\kappa^r_i(\tau )
\epsilon^{\mu}_r(u(p_s))\Big] =\sum_{i=1}^Np^{\mu}_i(\tau )=p^{\mu}_s,
\nonumber \\
&&{}\nonumber \\
 M_{sys}&=&P^{\mu}_T u_{\mu}(p_s)=-\lambda(\tau )
\sum_{i=1}^N {{m_i}\over {\sqrt{\lambda^2(\tau )- [{\dot {\vec
\eta}}_i(\tau )+\vec \lambda (\tau )]^2}}} \nonumber \\ &{\buildrel
\circ \over =}\,& {|}_{\lambda (\tau )=-1}\,\, \sum_{i=1}^N
\sqrt{m^2_i+{\vec \kappa}^2_i(\tau )},
\label{III12}
\end{eqnarray}

\noindent which turn out to be in the correct form only if
$\lambda (\tau )=-1$. This shows that the agreement with
parametrized Minkowski theories on arbitrary spacelike
hypersurfaces is obtained only on Wigner hyperplanes in the gauge
$T_s-\tau \approx 0$, which indeed implies $\lambda (\tau )=-1$.

\subsection{The Phase-Space Version of the Standard Energy-Momentum Tensor.}

The same result may be obtained by first reformulating the
standard energy-momentum in phase space and then by imposing the
restriction $p^{\mu}_i(\tau )=$\hfill\break
$\sqrt{m^2_i-\gamma^{rs}(\tau ,{\vec \eta} _i(\tau
))\kappa_{ir}(\tau )\kappa_{is}(\tau )} l^{\mu}(\tau ,{\vec \eta}
_i(\tau ))-\kappa_{ir}(\tau )\gamma^{rs}(\tau ,{\vec \eta}_i(\tau
)) z^{\mu}_s (\tau ,{\vec \eta}_i(\tau ))\, \rightarrow \,
\sqrt{m^2_i+{\vec \kappa}^2_i (\tau
)}u^{\mu}(p_s)+\kappa^r_i\epsilon^{\mu}_r(u(p_s))$ \footnote{In
this way we are sure to have imposed the restriction to positive
energy particles and  to have excluded the other $2^N-1$ branches
of the total mass spectrum.}[the last equality refers to the
Wigner hyperplane, see Eq.(\ref{II14})]:

\begin{eqnarray}
T^{\mu\nu}(z(\tau ,\vec \sigma ))&=&\sum_{i=1}^N{1\over {m_i}}
\int d\tau_1 \sqrt{{\dot x}^2_i(\tau_1)}p^{\mu}_i(\tau_1)p^{\nu}_i(\tau_1)
\delta^4(x_i(\tau_1)-z(\tau ,\vec \sigma ))=\nonumber \\
&=&\sum_{i=1}^N {{\sqrt{{\dot x}_i^2(\tau )}}\over
{m_i \sqrt{g(\tau ,\vec \sigma )}}}
p^{\mu}_i(\tau )p^{\nu}_i(\tau ) \delta^3(\vec \sigma -{\vec \eta}_i(\tau )),
\nonumber \\
\Downarrow && on\, Wigner's\, hyperplanes\nonumber \\
T^{\mu\nu}[x^{\beta}_s(\tau )+\epsilon^{\beta}_u(u(p_s))\sigma^u]&=&\sum_{i=1}
^N {{ \sqrt{{\dot x}^2_s+2{\dot x}_{s\beta}\epsilon^{\beta}_u(u(p_s))
{\dot \eta}^u_i-{\dot {\vec \eta}}^2_i}}\over {m_i \sqrt{{\dot x}_s\cdot
u(p_s)}}} (\tau ) p^{\mu}_i(\tau )p^{\nu}_i
(\tau )\delta^3(\vec \sigma -{\vec \eta}_i(\tau ))=\nonumber \\
&=&\sum_{i=1}^N \delta^3(\vec \sigma -{\vec \eta}_i(\tau ))
{{ \sqrt{{\dot x}^2_s+2{\dot x}_{s\beta}\epsilon^{\beta}_u(u(p_s))
{\dot \eta}^u_i-{\dot {\vec \eta}}_i^2}}\over {m_i \sqrt{{\dot x}_s\cdot
u(p_s)}}} (\tau )\nonumber \\
&&\Big[(m_i^2+{\vec \kappa}^2_i(\tau ))u^{\mu}(p_s)u^{\nu}(p_s)+\nonumber \\
&+&k^r_i(\tau )
\sqrt{m_i^2+{\vec \kappa}^2_i(\tau )}\Big(u^{\mu}(p_s)\epsilon^{\nu}_r(u(p_s))
+u^{\nu}(p_s)\epsilon^{\mu}_r(u(p_s))\Big) +\nonumber \\
&+&\kappa^r_i(\tau )\kappa^s_i(\tau )
\epsilon^{\mu}_r(u(p_s))\epsilon^{\nu}_s(u(p_s))\Big] =\nonumber \\
&=&\sum_{i=1}^N \delta^3(\vec \sigma -{\vec \eta}_i(\tau ))
{{ \sqrt{\lambda^2(\tau )-[{\dot {\vec \eta}}_i(\tau )+\vec \lambda (\tau )]^2
}}\over { \sqrt{-\lambda (\tau )}}} \nonumber \\
&&{{\sqrt{m_i^2+{\vec \kappa}^2_i(\tau )}}\over {m_i}}
\Big[\sqrt{m_i^2+{\vec \kappa}^2_i(\tau )}u^{\mu}(p_s)u^{\nu}(p_s)+\nonumber \\
&+&k^r_i(\tau )\Big(u^{\mu}(p_s)\epsilon^{\nu}_r(u(p_s))
+u^{\nu}(p_s)\epsilon^{\mu}_r(u(p_s))\Big) +\nonumber \\
&+&{{\kappa^r_i(\tau )\kappa^s_i(\tau )}\over {\sqrt{m_i^2+{\vec \kappa}^2
_i(\tau )}}}\epsilon^{\mu}_r(u(p_s))\epsilon^{\nu}_s(u(p_s))\Big] .
\label{III13}
\end{eqnarray}

Now the total 4-momentum and total mass are

\begin{eqnarray}
P^{\mu}_T&=&\int d^3\sigma T^{\mu\nu}[x^{\beta}_s(\tau )+\epsilon^{\beta}
_u(u(p_s))\sigma^u] u_{\nu}(p_s)=\nonumber \\
&=&\sum_{i=1}^N
{{ {\dot x}^2_s+2{\dot x}_{s\beta}\epsilon^{\beta}_u(u(p_s)){\dot \eta}^u_i-
{\dot {\vec \eta}}_i^2}\over { {\dot x}_s\cdot u(p_s)}} (\tau ) {{\sqrt{m^2_i+
{\vec \kappa}^2_i(\tau )} }\over {m_i}}\nonumber \\
&&[\sqrt{m^2_i+{\vec \kappa}^2_i(\tau )} u^{\mu}(p_s)+\kappa^r_i(\tau )
\epsilon^{\mu}_r(u(p_s))]=\nonumber \\
&=&\sum_{i=1}^N {{ \sqrt{\lambda^2(\tau )-[{\dot {\vec \eta}}_i(\tau
)+\vec \lambda (\tau )]^2 }}\over { \sqrt{-\lambda (\tau
)}}}{{\sqrt{m^2_i+ {\vec \kappa}^2_i(\tau )} }\over {m_i}}
p^{\mu}_i(\tau ),\nonumber \\
 M_{sys}&=&P^{\mu}_T u_{\mu}(p_s)=\sum_{i=1}^N
{{ \sqrt{\lambda^2(\tau )-[{\dot {\vec \eta}}_i(\tau )+\vec \lambda
(\tau )]^2 }}\over { {-\lambda (\tau )}}} {{\sqrt{m^2_i+ {\vec
\kappa}^2_i(\tau )} }\over {m_i}} \sqrt{m^2_i+{\vec \kappa}^2_i(\tau
)}.
\label{III14}
\end{eqnarray}

These equations show that the total 4-momentum evaluated from the
energy-momentum tensor of the standard theory restricted to
positive energy particles is consistent \footnote{Namely we get
$P^{\mu}_T=p^{\mu}_s$ and $M_{sys}=\sum_{i=1}^N\sqrt{m^2_i+{\vec
\kappa}^2_i(\tau )}$.} with the description on the Wigner
hyperplanes with its gauge freedom $\lambda (\tau )$, $\vec
\lambda (\tau )$, {\it only} by working with the Dirac brackets of
the gauge $T_s\equiv \tau$, where one has $\lambda (\tau )=-1$ and

\begin{eqnarray}
{\dot x}^{\mu}_s(\tau )&=&u^{\mu}(p_s)+\epsilon^{\mu}_r(u(p_s))
\lambda_r(\tau ),\nonumber \\
x^{\mu}_s(\tau )&=&x^{\mu}_s(0)+\tau u^{\mu}(p_s)+\epsilon^{\mu}_r
(u(p_s)) \int^{\tau}_0d\tau_1 \lambda_r(\tau_1),
\label{III15}
\end{eqnarray}

\noindent because $m_i/\sqrt{1- [{\dot {\vec \eta}}_i(\tau )+\vec \lambda (\tau
)]^2}\, {\buildrel \circ \over
=}\, \sqrt{m^2_ic^2+{\vec \kappa}^2_i(\tau )}$.

Therefore,  for every $\vec \lambda (\tau )$, we get

\begin{eqnarray}
T^{\mu\nu}[x^{\beta}_s(T_s)+\epsilon^{\beta}_u(u(p_s))\sigma^u]&=&
\epsilon^{\mu}_A(u(p_s)) \epsilon^{\nu}_B(u(p_s)) T^{AB}(T_s,\vec \sigma )
=\nonumber \\
 &=&\sum_{i=1}^N \delta^3(\vec \sigma -{\vec \eta}_i(T_s))
\Big[\sqrt{m_i^2+{\vec \kappa}^2_i(T_s)}
u^{\mu}(p_s)u^{\nu}(p_s)+\nonumber \\
&+&k^r_i(T_s)\Big(u^{\mu}(p_s)\epsilon^{\nu}_r(u(p_s))
+u^{\nu}(p_s)\epsilon^{\mu}_r(u(p_s))\Big) +\nonumber \\
&+&{{\kappa^r_i(T_s)\kappa^s_i(T_s)}\over {\sqrt{m_i^2+{\vec \kappa}^2
_i(T_s)} }} \epsilon^{\mu}_r(u(p_s))\epsilon^{\nu}_s(u(p_s))\Big] ,
\nonumber \\
T^{\tau\tau}(T_s,\vec \sigma )&=&\sum_{i=1}^N\delta^3(\vec \sigma
-{\vec \eta}_i(T_s)) \sqrt{m_i^2+{\vec \kappa}_i^2(T_s))},\nonumber \\
T^{\tau r}(T_s,\vec \sigma )&=&\sum_{i=1}^N\delta^3(\vec \sigma
-{\vec \eta}_i(T_s)) \kappa_i^r(T_s),\nonumber \\
T^{rs}(T_s,\vec \sigma )&=&\sum_{i=1}^N\delta^3(\vec \sigma
-{\vec \eta}_i(T_s)) {{ \kappa_i^r(T_s) \kappa_i^s(T_s)}\over
{ \sqrt{m_i^2+{\vec \kappa}_i^2(T_s)}}},\nonumber \\
 &&{}\nonumber \\
P^{\mu}_T&=&p^{\mu}_s=Mu^{\mu}(p_s)+\epsilon^{\mu}_r(u(p_s))\kappa^r_{+}
\approx M u^{\mu}(p_s),\nonumber \\
M&=&\sum_{i=1}^N \sqrt{m^2_ic^2+{\vec \kappa}_i^2(T_s)},\nonumber \\
 &&{}\nonumber \\
T^{\mu\nu}[x^{\beta}_s(T_s)+\epsilon^{\beta}_u(u(p_s))\sigma^u]
&&u_{\nu}(p_s)=\epsilon^{\mu}_A(u(p_s)) T^{A\tau}(T_s,\vec \sigma )=
\nonumber \\
 &=&\sum_{i=1}^N \delta^3(\vec \sigma -{\vec \eta}_i(T_s))
\nonumber \\
&&[\sqrt{m^2_ic^2+{\vec \kappa}_i^2(T_s)}u^{\mu}(p_s)+\kappa^r_i(T_s)\epsilon
^{\mu}_r(u(p_s))],\nonumber \\
T^{\mu}{}_{\mu}[x^{\beta}_s(T_s)
+\epsilon^{\beta}_u(u(p_s))\sigma^u]&=& T^A{}_A(T_s,\vec \sigma
)=\nonumber \\
 &=&\sum_{i=1}^N \delta^3(\vec \sigma -{\vec \eta}_i(T_s)) {{m^2_i}\over
{\sqrt{m^2_i+{\vec \kappa}^2_i(T_s)}}}.
\label{III16}
\end{eqnarray}

\vfill\eject

\section{Dixon's Multipoles on the Wigner Hyperplane.}

In this Section we  shall define the special relativistic Dixon
multipoles on the Wigner hyperplane for the N-body
problem\footnote{Previous studies of multipole theories of point
particles can be found in
Refs.\cite{mul4,mul8,mul2,mul10,mul9,mul11,mul12,mul13}.}.  We
list the non-relativistic multipoles for N free particles in
Appendix A by comparison.

Let us now consider an arbitrary timelike worldline $w^{\mu}(\tau
)=z^{\mu}(\tau ,\vec \eta (\tau ))=x_s^{({\vec q}_{+})\mu}(\tau )+
\epsilon^{\mu}_r(u(p_s))\eta^r(\tau )$ and let us evaluate
the Dixon multipoles \cite{dixon} \footnote{See this paper for the
previous definitions given by Bielecki, Mathisson and
Weyssenhof\cite{mul1,mul2,mul3,mul4}.} on the Wigner hyperplanes in
the natural gauge with respect to the given worldline. A generic point
will be parametrized as

\begin{equation}
z^{\mu}(\tau ,\vec \sigma )=x_s^{({\vec q}_{+})\mu}(\tau
)+\epsilon^{\mu}_r(u(p_s))\sigma^r=w^{\mu}(\tau )+
\epsilon^{\mu}_r(u(p_s)) [\sigma^r-\eta^r(\tau )]\, {\buildrel
{def} \over =}\, w^{\mu}(\tau )+\delta z^{\mu}(\tau ,\vec \sigma ),
\label{IV1}
\end{equation}

\noindent so that $\delta z_{\mu}(\tau ,\vec \sigma )u^{\mu}(p_s)=0$.

For $\vec \eta (\tau )=0$ we will get the multipoles with respect to
the origin $x_s^{({\vec q}_{+})\mu}(\tau )$.

\subsection{Dixon's Multipoles.}

Lorentz covariant {\it Dixon's multipoles} and their Wigner
covariant counterparts on the Wigner hyperplanes are defined as

\begin{eqnarray}
t_T^{\mu_1...\mu_n\mu\nu}(T_s,\vec
\eta )&=&t_T^{(\mu_1...\mu_n)(\mu\nu)}(T_s,\vec \eta )=
\nonumber \\
 &=&\epsilon^{\mu_1}_{r_1}(u(p_s))...\epsilon^{\mu_n}_{r_n}(u(p_s))\,
 \epsilon^{\mu}_A(u(p_s))\epsilon^{\nu}_B(u(p_s)) q_T^{r_1..r_nAB}(T_s,\vec \eta )
 =\nonumber \\
 &&{}\nonumber \\
&=&\int d^3\sigma \delta z^{\mu_1}(T_s,\vec \sigma )...\delta
z^{\mu_n}(T_s,\vec
\sigma ) T^{\mu\nu}[x^{({\vec q}_{+})\beta}_s(T_s)+\epsilon^{\beta}_u(u(p_s))
\sigma^u]=\nonumber \\
&=&\epsilon^{\mu}_A(u(p_s))\epsilon^{\nu}_B(u(p_s)) \int d^3\sigma
\delta z^{\mu_1}(T_s,\vec \sigma )....\delta z^{\mu_n}(T_s,\vec \sigma )
T^{AB}(T_s,\vec \sigma )=\nonumber \\
 &&{}\nonumber \\
&=&\epsilon^{\mu_1}_{r_1}(u(p_s))...\epsilon^{\mu_n}_{r_n}(u(p_s))
\nonumber \\
 &&\Big[ u^{\mu}(p_s) u^{\nu}(p_s)
\sum_{i=1}^N[\eta^{r_1}_i(T_s)-\eta^{r_1}(T_s)]...[\eta^{r_n}
_i(T_s)-\eta^{r_n}(T_s)]\sqrt{m^2_i+{\vec \kappa}^2_i(T_s)}+\nonumber \\
 &+&\epsilon^{\mu}_r(u(p_s))\epsilon^{\nu}_s(u(p_s))\nonumber \\
 &&\sum_{i=1}^N[\eta^{r_1}
_i(T_s)-\eta^{r_1}(T_s)]...[\eta^{r_n}_i(T_s)-\eta^{r_n}(T_s)]
{{\kappa^r_i(T_s)\kappa^s_i(T_s)}\over
{\sqrt{m_i^2+{\vec \kappa}^2_i(T_s)}}}+\nonumber \\
&+&[u^{\mu}(p_s)\epsilon^{\nu}_r(u(p_s))+u^{\nu}(p_s)\epsilon^{\mu}_r(u(p_s))]
  \nonumber \\
&&\sum_{i=1}^N[\eta^{r_1}_i(T_s)-\eta^{r_1}(T_s)]...[\eta^{r_n}_i(T_s)-
\eta^{r_n}(T_s)]\kappa^r_i(T_s)\Big] ,\nonumber \\
 &&{}\nonumber \\
 q_T^{r_1...r_nAB}(T_s,\vec \eta )&=&\delta^A_{\tau}\delta^B_{\tau}
 \sum_{i=1}^N[\eta^{r_1}_i(T_s)-\eta^{r_1}(T_s)]...[\eta^{r_n}
_i(T_s)-\eta^{r_n}(T_s)]\sqrt{m^2_i+{\vec \kappa}^2_i(T_s)}+\nonumber \\
&+&\delta^A_u\delta^B_v \sum_{i=1}^N[\eta^{r_1}
_i(T_s)-\eta^{r_1}(T_s)]...[\eta^{r_n}_i(T_s)-\eta^{r_n}(T_s)]
{{\kappa^u_i(T_s)\kappa^v_i(T_s)}\over {\sqrt{m_i^2+{\vec
\kappa}^2_i(T_s)}}}+\nonumber \\
&+&(\delta^A_{\tau}\delta^B_u+\delta^A_u\delta^B_{\tau})
\sum_{i=1}^N[\eta^{r_1}_i(T_s)-\eta^{r_1}(T_s)]...[\eta^{r_n}_i(T_s)-
\eta^{r_n}(T_s)]\kappa^r_i(T_s),\nonumber \\
 &&{}\nonumber \\
u_{\mu_1}(p_s)&& t_T^{\mu_1...\mu_n\mu\nu}(T_s,\vec \eta )=0,
\nonumber \\
 &&{}\nonumber \\
 t_T^{\mu_1...\mu_n\mu}{}_{\mu}(T_s,\vec \eta )
 &{\buildrel {def}
\over =}&\epsilon^{\mu_1}_{r_1}(u(p_s))...\epsilon^{\mu_n}
_{r_n}(u(p_s))q_T^{r_1...r_nA}{}_A(T_s,\vec \eta )=\nonumber \\
 &=&\int d^3\sigma
\delta z^{\mu_1}(\tau ,\vec
\sigma )...\delta z^{\mu_n}(\tau ,\vec \sigma ) T^{\mu}{}_{\mu}
[x^{({\vec q}_{+})\mu}_s(T_s)+\epsilon^{\mu}_u(u(p_s))\sigma^u]
=\nonumber \\
&=&\epsilon^{\mu_1}_{r_1}(u(p_s))...\epsilon^{\mu_n}_{r_n}(u(p_s))
\nonumber \\
 &&\sum_{i=1}^N[\eta^{r_1}_i(T_s)-\eta^{r_1}(T_s)]...
[\eta^{r_n}_i(T_s)-\eta^{r_n}(T_s)] {{m^2_i}\over {\sqrt{m^2_i+ {\vec
\kappa}_i^2(T_s)}}}=\nonumber \\
 &&{}\nonumber \\
 {\tilde t}_T^{\mu_1...\mu_n}(T_s,\vec \eta )&=&
 t_T^{\mu_1...\mu_n\mu\nu}(T_s,\vec \eta )u_{\mu}(p_s)
u_{\nu}(p_s)=\nonumber \\
&=&\epsilon^{\mu_1}_{r_1}(u(p_s))...\epsilon^{\mu_n}_{r_n}(u(p_s))
q_T^{r_1...r_n\tau \tau}(T_s,\vec \eta )=\nonumber \\
&=&\epsilon^{\mu_1}_{r_1}(u(p_s))...\epsilon^{\mu_n}_{r_n}(u(p_s))
\nonumber \\
 &&\sum_{i=1}^N[\eta^{r_1}_i(T_s)-\eta^{r_1}(T_s)]...[\eta^{r_n}
_i(T_s)-\eta^{r_n}(T_s)]\sqrt{m^2_i+{\vec \kappa}^2_i(T_s)}.
\label{IV2}
\end{eqnarray}

\noindent  Related multipoles are

\begin{eqnarray}
p_T^{\mu_1..\mu_n\mu}(T_s,\vec \eta )&=&
t_T^{\mu_1...\mu_n\mu\nu}(T_s,\vec \eta ) u_{\nu}(p_s)=
\nonumber \\
 &=&\epsilon^{\mu_1}_{r_1}(u(p_s))...\epsilon^{\mu_n}_{r_n}(u(p_s))
 \epsilon^{\mu}_A(u(p_s)) q_T^{r_1...r_nA\tau}(T_s,\vec \eta )=\nonumber \\
 &=&\epsilon^{\mu_1}_{r_1}(u(p_s)).... \epsilon^{\mu_n}_{r_n}(u(p_s))
   \nonumber \\
 &&\sum_{i=1}^N [\eta^{r_1}_i(T_s)-\eta^{r_1}(T_s)]...[\eta^{r_n}_i(T_s)-
 \eta^{r_n}(T_s)]\nonumber \\
&&\Big[\sqrt{m_i^2+{\vec \kappa}^2_i(\tau )}u^{\mu}(p_s)+ k^r_i(T_s)
\epsilon^{\mu}_r(u(p_s))\Big] ,\nonumber \\
 &&{}\nonumber \\
&&u_{\mu_1}(p_s) p_T^{\mu_1...\mu_n\mu}(T_s,\vec \eta )=0,\nonumber \\
&&p_T^{\mu_1...\mu_n\mu}(T_s,\vec \eta )u_{\mu}(p_s)={\tilde
t}_T^{\mu_1...\mu_n}(T_s,\vec \eta ),\nonumber \\
 &&{}\nonumber \\
n=0&&\Rightarrow p^{\mu}_T(T_s,\vec \eta )=\epsilon^{\mu}_A(u(p_s))
q_T^{A\tau}(T_s)=P^{\mu}_T\approx p^{\mu}_s.
\label{IV3}
\end{eqnarray}

The inverse formulas, giving the {\it multipolar expansion},  are

\begin{eqnarray}
T^{\mu\nu}[w^{\beta}(T_s)+\delta z^{\beta}(T_s,\vec \sigma )]&=&
T^{\mu\nu}[x^{({\vec
q}_{+})\beta}_s(T_s)+\epsilon^{\beta}_r(u(p_s))\sigma^r]=\nonumber \\
 &=&\epsilon^{\mu}_A(u(p_s))\epsilon^{\nu}_B(u(p_s)) T^{AB}(T_s,\vec \sigma )
 =\nonumber \\
  &&{}\nonumber \\
  &=&\epsilon^{\mu}_A(u(p_s))\epsilon^{\nu}_B(u(p_s)) \sum_{n=0}^{\infty}
  (-1)^n\, {{ q_T^{r_1...r_nAB}(T_s,\vec \eta )}\over {n!}} \nonumber \\
  &&{{\partial^n}\over
  {\partial \sigma^{r_1}...\partial \sigma^{r_n}}} \delta^3(\vec \sigma -
  \vec \eta (T_s))=\nonumber \\
  &=&\sum_{n=0}^{\infty} (-1)^n\, {{t_T^{\mu_1...\mu_n\mu\nu}(T_s,\vec \eta )}\over
  {n!}} \epsilon_{r_1\mu_1}(u(p_s))...\epsilon_{r_n\mu_n}(u(p_s))
  \nonumber \\
  &&{{\partial^n}\over
  {\partial \sigma^{r_1}...\partial \sigma^{r_n}}} \delta^3(\vec \sigma -
  \vec \eta (T_s)).
\label{IV4}
\end{eqnarray}

The quantities $q_T^{r_1...r_n\tau\tau}(T_s,\vec \eta )$,
$q_T^{r_1...r_n r\tau}(T_s,\vec \eta )=q_T^{r_1...r_n \tau r}(T_s,\vec
\eta )$, $q_T^{r_1...r_n uv}(T_s,\vec \eta )$ are the {\it mass density, stress tensor and
momentum density multipoles} with respect to the worldline
$w^{\mu}(T_s)$  (barycentric for $\vec \eta =0$) respectively.

\subsection{Monopoles.}

The {\it monopoles} correspond to $n=0$
\footnote{$t_T^{\mu\nu}(T_s,\vec \eta
)=\epsilon^{\mu}_A(u(p_s))\epsilon^{\nu}_B(u(p_s))q_T^{AB}(T_s,\vec
\eta )$, $p_T^{\mu}(T_s,\vec \eta
)=\epsilon^{\mu}_A(u(p_s))q^{A\tau}_T(T_s,\vec \eta )$, ${\tilde
t}_T(T_s,\vec \eta )=q_T^{\tau \tau}(T_s,\vec \eta )$.} and have
the following expression \footnote{They are $\vec \eta$
independent; see Appendix C of Ref.\cite{iten1} and Appendix A for
the non-relativistic limit.} \footnote{In this Section we use
results from Section V of Ref.\cite{iten1}, where the rest-frame
(${\vec \kappa}_{+}={\vec q}_{+}=0$) canonical relative variables
with respect to the {\it internal} 3-center of mass were found by
means of a classical Gartenhaus-Schwartz
transformation\cite{garten}. This transformation is a sequence of
canonical transformations depending on a parameter $\alpha$. The
final rest-frame relative variables are obtained in the limit
$\alpha \rightarrow \infty$. In Eqs.(\ref{IV5}), (\ref{IV9}),
(\ref{IV11}), (\ref{IV13}), (\ref{IV14}) we use the following
symbols from Ref.\cite{iten1}: ${\vec \kappa}_i(\infty )=\sqrt{N}
\sum_{a=1}^{N-1}\gamma_{ai}{\vec \pi}_{qa}$, $\Pi =M^2_{sys}-{\vec
\kappa}_{+}^2$, $H_i=\sqrt{m^2_i+{\vec \kappa}^2_i}$, $H_i(\infty
)= \sqrt{m^2_i+N \sum_{ab}^{1..N-1} \gamma_{ai}\gamma_{bi} {\vec
\pi}_{qa}\cdot {\vec \pi}_{qb}}$. Also the non-relativistic
limits, $c \rightarrow \infty$, are shown. }

\begin{eqnarray}
q^{AB}_T(T_s,\vec \eta )&=&\delta^A_{\tau}\delta^B_{\tau} M
+\delta^A_u\delta^B_v\sum_{i=1}^N{{\kappa^u_i\kappa^v_i}\over
{\sqrt{m_i^2+{\vec \kappa}_i^2}}}+
(\delta^A_{\tau}\delta^B_u+\delta^A_u\delta^B_{\tau}) \kappa^u_{+}
\approx \nonumber \\
&{\rightarrow}_{\alpha \rightarrow \infty}\,&
\delta^A_{\tau}\delta^B_{\tau} \sqrt{\Pi} +\delta^A_u\delta^B_v \sum_{i=1}^N
{{\kappa_i^u(\infty )\kappa_i^v(\infty )}\over {H_i(\infty
)}}=\nonumber \\
 &=&\delta^A_{\tau}\delta^B_{\tau}  \sum_{i=1}^N
 \sqrt{m_i^2+N\sum_{de}\gamma_{di}\gamma_{ei}{\vec \pi}_{qd}\cdot {\vec \pi}_{qe}}
+\nonumber \\
 &+&\delta^A_u\delta^B_v N \sum_{i=1}^N {{ \sum_{ab}^{1..N-1}\gamma_{ai}
 \gamma_{bi} {\vec \pi}_{qa}\cdot{\vec \pi}_{qb}}\over
 {\sqrt{m_i^2+N\sum_{de}\gamma_{di}\gamma_{ei}{\vec \pi}_{qd}\cdot {\vec \pi}_{qe}}}},
 \nonumber \\
 &&{}\nonumber \\
 q^{\tau\tau}_T(T_s,\vec \eta )\, &{\rightarrow}_{c \rightarrow \infty}\,& \sum_{i=1}^N
 m_ic^2 +{1\over 2}\sum_{ab}^{1..N-1}\sum_{i=1}^N{{N\gamma_{ai}\gamma_{bi}}\over {m_i}}
 {\vec \pi}_{qa}\cdot {\vec \pi}_{qb} +O(1/c)=\nonumber \\
 &=&\sum_{i=1}^Nm_ic^2+ H_{rel,nr} +O(1/c),\nonumber \\
 q_T^{r\tau}(T_s,\vec \eta ) &=&\kappa^r_{+} \approx 0,\quad rest-frame\, condition\,
 (also\, at\, the \, non-relativistic\, level),\nonumber \\
 q^{uv}_T(T_s,\vec \eta )\, &{\rightarrow}_{c \rightarrow \infty}\,& \sum_{ab}^{1..N-1}
 \sum_{i=1}^N {{N\gamma_{ai}\gamma_{bi}}\over {m_i}} \pi^u_{qa}\pi^v_{qb} +O(1/c)
 =\nonumber \\
 &=& \sum_{ab}^{1..N-1} k^{-1}_{ab} \pi^u_{qa}\pi^v_{qb} +O(1/c) =\sum_{ab}^{1..N-1}
 k_{ab} {\dot \rho}^u_a{\dot \rho}^v_b +O(1/c), \nonumber \\
 &&{}\nonumber \\
 q^A_{T A}(T_s,\vec \eta )&=& t^{\mu}_{T \mu}(T_s,\vec \eta )=
 \sum_{i=1}^N {{m^2_i}\over {\sqrt{m_i^2+
 {\vec \kappa}_i^2}}}\nonumber \\
 &{\rightarrow}_{\alpha \rightarrow \infty}\,& \sum_{i=1}^N {{m_i^2}\over {H_i(\infty )}}=
 \sum_{i=1}^N{{m_i^2}\over
 {\sqrt{m_i^2+N\sum_{de}\gamma_{di}\gamma_{ei}{\vec \pi}_{qd}\cdot {\vec \pi}_{qe}}}}
 \nonumber \\
 &{\rightarrow}_{c \rightarrow \infty}\,& \sum_{i=1}^N m_ic^2 -{1\over 2}\sum_{ab}^{1..N-1}
 \sum_{i=1}^N {{N\gamma_{ai} \gamma_{bi}}\over {m_i}} {\vec \pi}_{qa}\cdot {\vec \pi}_{qb} +O(1/c)
 =\nonumber \\
 &=&\sum_{i=1}^Nm_ic^2 -H_{rel,nr} +O(1/c).
\label{IV5}
\end{eqnarray}

\noindent where we used Eqs. (5.10), (5.11) of Ref.\cite{iten1} to
obtain their expression in terms of the internal relative
variables.

Therefore, in the rest-frame instant form, the {\it mass monopole}
is the invariant mass $M=\sum_{i=1}^N\sqrt{m_i^2+{\vec
\kappa}_i^2}$, while the {\it momentum monopole} vanishes.

\subsection{Dipoles.}

The {\it dipoles} correspond to $n=1$
\footnote{$t_T^{\mu_1\mu\nu}(T_s,\vec
\eta
)=\epsilon^{\mu_1}_{r_1}(u(p_s))$$\epsilon^{\mu}_A(u(p_s))\epsilon^{\nu}_B(u(p_s))
q_T^{r_1AB}(T_s,\vec \eta )$, ${\tilde t}^{\mu_1}_T(T_s,\vec \eta
)=\epsilon^{\mu_1}_{r_1}(u(p_s))q_T^{r_1\tau\tau}(T_s,\vec \eta )$,
$p_T^{\mu_1\mu}(T_s,\vec \eta )=\epsilon^{\mu_1}_{r_1}(u(p_s))
\epsilon^{\mu}_A(u(p_s))q_T^{r_1A\tau}(T_s,\vec \eta )$.}:

\begin{eqnarray}
q_T^{rAB}(T_s,\vec \eta )&=&\delta^A_{\tau}\delta^B_{\tau} M
[R^r_{+}(T_s)-\eta^r(T_s)] +\delta^A_u\delta^B_v\Big[
\sum_{i=1}^N{{\eta_i^r\kappa^u_i\kappa^v_i}\over {\sqrt{m_i^2+
{\vec \kappa}_i^2}}}(T_s)-\eta^r(T_s)q_T^{uv}(T_s,\vec \eta )\Big] +
\nonumber \\
&+&(\delta^A_{\tau}\delta^B_u+\delta^A_u\delta^B_{\tau}) \Big[
\sum_{i=1}^N[\eta^r_i\kappa_i^u](T_s)-\eta^r(T_s)\kappa^u_{+}\Big] ,
\nonumber \\
 &&{}\nonumber \\
 q_T^{rA}{}_A(T_s,\vec \eta )&=&\epsilon^{r_1}_{\mu_1}(u(p_s))
 t_T^{\mu_1\mu}{}_{\mu}(T_s,\vec \eta )=\sum_{i=1}^N {{[\eta^r_i-\eta^r]
 m_i^2}\over {\sqrt{m_i^2+{\vec \kappa}_i^2}}}(T_s).
\label{IV6}
\end{eqnarray}

The vanishing of the {\it mass dipole} identifies the {\it
internal M$\o$ller center of energy} ${\vec R}_{+}\approx {\vec
q}_{+}\approx {\vec y}_{+}$ and therefore the {\it rest-frame
internal center of mass} ${\vec q}_{+}$:

\begin{equation}
q_T^{r\tau\tau}(T_s,\vec \eta )=\epsilon^{r_1}_{\mu_1}(u(p_s))
{\tilde t}_T^{\mu_1}(T_s,\vec \eta )=0,\quad \Rightarrow \vec \eta
(T_s)={\vec R}_{+} [=0\,\,\, for\,\,\, x^{\mu}_s=x_s^{({\vec
q}_{+}) \mu}]. \label{IV7}
\end{equation}

The time derivative of the mass dipole identifies the {\it
center-of-mass momentum-velocity relation} for the  system when $\vec
\eta
=0$

\begin{equation}
{{d q_T^{r\tau\tau}(T_s,\vec \eta )}\over {dT_s}}\, {\buildrel
\circ \over =}\, \kappa^r_{+}-M {\dot \eta}^r(T_s)\,
{\rightarrow}_{\vec \eta \rightarrow 0}\, 0.
\label{IV8}
\end{equation}

The expression of the dipoles in terms of the internal relative
variables when $\vec \eta ={\vec R}_{+}={\vec q}_{+}=0$ is obtained by
using Eqs. (5.10), (5.11), (5.22), (5.25) of Ref.\cite{iten1}

\begin{eqnarray}
q_T^{r\tau\tau}(T_s,{\vec R}_{+})&=&0,\nonumber \\
 q_T^{ru\tau}(T_s,{\vec R}_{+})&=&\sum_{i=1}^N\eta_i^r\kappa_i^u-R_{+}^r\kappa_{+}^u=
 \sum_{a=1}^{N-1} \rho_a^r\pi^u_a+(\eta^r_{+}-R^r_{+})\kappa^u_{+}\nonumber \\
 &{\rightarrow}_{\alpha \rightarrow \infty}\,& \sum_{a=1}^{N-1} \rho^r_{qa}
 \pi^u_{qa}\nonumber \\
 &{\rightarrow}_{c \rightarrow \infty}\,& \sum_{a=1}^{N-1} \rho^r_a\pi^u_{qa} =
 \sum{ab}^{1..N-1} k_{ab} \rho^r_a{\dot \rho}^u_b,\nonumber \\
 q_T^{ruv}(T_s,{\vec R}_{+})&=& \sum_{i=1}^N \eta^r_i {{\kappa_i^u\kappa_i^v}\over {H_i}}-
 R^r_{+} \sum_{i=1}^N {{\kappa_i^u\kappa_i^v}\over {H_i}}=\nonumber \\
 &=&{1\over {\sqrt{N}}} \sum_{i=1}^N \sum_{a=1}^{N-1} \gamma_{ai} \rho^r_a
 {{\kappa^u_i\kappa_i^v}\over {H_i}}+(\eta^r_{+}-R^r_{+})\sum_{i=1}^N
 {{\kappa_i^u\kappa_i^v}\over {H_i}}\nonumber \\
 &{\rightarrow}_{\alpha \rightarrow \infty}\,& {1\over {\sqrt{N}}} \sum_{ij}^{1..N}
 \sum_{a=1}^{N-1} (\gamma_{ai}-\gamma_{aj}) \rho^r_{qa}
 {{H_j(\infty )\kappa_i^u(\infty )\kappa_i^v(\infty )}\over {\sqrt{\Pi}
 H_i(\infty )}}=\nonumber \\
 &=& \sum_{a=1}^{N-1} \Big( c\sqrt{N}\sum_{ij}^{1..N}(\gamma_{ai}-\gamma_{aj})
{{ \sqrt{m_j^2+N\sum_{de}\gamma_{dj}\gamma_{ej}{\vec \pi}_{qd}\cdot
{\vec \pi}_{qe}} }\over
{\sqrt{m_i^2+N\sum_{de}\gamma_{di}\gamma_{ei}{\vec \pi}_{qd}\cdot
{\vec \pi}_{qe}}}}  \times \nonumber \\
 &&{{ \sum_{bc}^{1..N-1} \gamma_{bi}\gamma_{ci} \pi_{qb}^u \pi^v_{qc}}\over
 {\sum_{k=1}^N\sqrt{m_k^2+N\sum_{de}\gamma_{dk}\gamma_{ek}
 {\vec \pi}_{qd}\cdot {\vec \pi}_{qe}} }} \Big) \rho^r_{qa}\nonumber \\
&{\rightarrow}_{c \rightarrow
\infty}\,&\sum_{ij}^{1..N}\sum_{a=1}^{N-1} {{\gamma_{ai}-\gamma_{aj}}\over {\sqrt{N}}}
\rho^r_a {{m_jN}\over {m_im}} \sum_{bc}^{1..N-1}\gamma_{bi}\gamma_{ci}\pi^u_{qb}\pi^v_{qc}+O(1/c)
=\nonumber \\
&=&{1\over {\sqrt{N}}}\sum_{abc}^{1..N-1}\Big[
N\sum_{i=1}^N{{\gamma_{ai}\gamma_{bi}\gamma_{ci}}\over
{m_i}}-{{\sum_{j=1}^Nm_j\gamma_{aj}}\over m}\Big]
\rho^r_a\pi^u_{qb}\pi^v_{qc}+O(1/c),\nonumber \\
 &&{}\nonumber \\
 q_T^{rA}{}_A(T_s,{\vec R}_{+})&=& \sum_{i=1}^N (\eta_i^r-R^r_{+}){{m_i^2}\over
 {H_i}}={1\over {\sqrt{N}}} \sum_{ij}^{1..N}\sum_{a=1}^{N-1}(\gamma_{ai}-\gamma_{aj})
 \rho^r_a {{m_iH_j}\over {H_iH_M}}\nonumber \\
 &{\rightarrow}_{\alpha \rightarrow \infty}\,& \sum_{a=1}^{N-1}\Big( {1\over
 {\sqrt{N}}} \sum_{ij}^{1..N}(\gamma_{ai}-\gamma_{aj})
 {{\sqrt{m_j^2+N\sum_{de}\gamma_{dj}\gamma_{ej}{\vec \pi}_{qd}\cdot {\vec \pi}_{qe}}}\over
 {\sqrt{m_i^2+N\sum_{de}\gamma_{di}\gamma_{ei}{\vec \pi}_{qd}\cdot {\vec \pi}_{qe}}}}
 \times \nonumber \\
 &&{{m_i}\over {\sum_{k=1}^N
 \sqrt{m_k^2+N\sum_{de}\gamma_{dk}\gamma_{ek}{\vec \pi}_{qd}\cdot {\vec \pi}_{qe}}}}
 \Big) \rho^r_a\nonumber \\
 &{\rightarrow}_{c \rightarrow \infty}\,&\sum_{ij}^{1..N}\sum_{a=1}^{N-1}\sqrt{N}
 (\gamma_{ai}-\gamma_{aj}) \rho^r_a {{m_jN}\over {m_im}}\sum_{bc}^{1..N-1}\gamma_{bi}\gamma_{ci}
 {\vec \pi}_{qb}\cdot {\vec \pi}_{qc}+O(1/c)=\nonumber \\
 &=&\sqrt{N}\sum_{abc}^{1..N-1}\Big[
N\sum_{i=1}^N{{\gamma_{ai}\gamma_{bi}\gamma_{ci}}\over
{m_i}}-{{\sum_{j=1}^Nm_j\gamma_{aj}}\over m}\Big] \rho^r_a {\vec
\pi}_{qb}\cdot {\vec \pi}_{qc} +O(1/c).
\label{IV9}
\end{eqnarray}

The antisymmetric part of the related dipole $p_T^{\mu_1\mu}(T_s,\vec
\eta )$ identifies the {\it spin tensor}. Indeed, the {\it spin dipole} \footnote{See
Ref.\cite{dixon} for the ambigous definitions of
$p_T^{\mu}$, $S_T^{\mu\nu}$ given by Papapetrou, Urich and
 Papapetrou, and by B.Tulczyjev and W.Tulczyjev\cite{mul5,mul6,mul7}.} is

\begin{eqnarray}
S^{\mu\nu}_T(T_s)[\vec \eta]&=&2 p_T^{[\mu\nu ]}(T_s,\vec \eta )=
\nonumber \\
 &=&M [R^r_{+}(T_s)-\eta^r(T_s)]
\Big[ \epsilon^{\mu}_r(u(p_s))u^{\nu}(p_s)-\epsilon^{\nu}_r(u(p_s))u^{\mu}(p_s)
\Big] +\nonumber \\
&+&\sum_{i=1}^N[\eta^r_i(T_s)-\eta^r(T_s)]\kappa^s_i(T_s)
\Big[\epsilon^{\mu}_r(u(p_s))
\epsilon^{\nu}_s(u(p_s))-\epsilon^{\nu}_r(u(p_s))\epsilon^{\mu}_s(u(p_s))
\Big],\nonumber \\
 &&{}\nonumber \\
 &&\Downarrow \nonumber \\
 &&{}\nonumber \\
S^{\mu\nu}_T(T_s)[\vec \eta =0]&=&S^{\mu\nu}_s\, {\buildrel \circ
\over =}\nonumber \\
 &{\buildrel \circ \over =}\,&
\sum_{i=1}^N{{m_i\eta^r_i(T_s)}\over {\sqrt{1-{\dot {\vec \eta}}
_i^2(T_s)}}}\Big[ \epsilon^{\mu}_r(u(p_s))u^{\nu}(p_s)-\epsilon^{\nu}
_r(u(p_s))u^{\mu}(p_s)\Big] +\nonumber \\
 &+&\sum_{i=1}^N{{m_i\eta^r_i(T_s){\dot \eta}^s_i(T_s)}\over
{\sqrt{1-{\dot {\vec
\eta}}_i^2(T_s)}}}\Big[\epsilon^{\mu}_r(u(p_s))
\epsilon^{\nu}_s(u(p_s))-\epsilon^{\nu}_r(u(p_s))\epsilon^{\mu}_s(u(p_s))
\Big] ,\nonumber \\
 &&{}\nonumber \\
m^{\nu}_{u(p_s)}(T_s,\vec \eta )&=& u_{\mu}(p_s)
S^{\mu\nu}_T(T_s)[\vec
\eta ]=-\epsilon^{\nu}_r(u(p_s)) [{\bar S}^{\tau r}_s-M\eta^r(T_s)]=
\nonumber \\
&=&-\epsilon^{\nu}_r(u(p_s)) M[R^r_{+}(T_s)-\eta^r(T_s)]=-
\epsilon^{\nu}_r(u(p_s)) q_T^{r\tau\tau}(T_s,\vec \eta ),\nonumber \\
 &&{}\nonumber \\
 \Rightarrow&& u_{\mu}(p_s) S^{\mu\nu}_T(T_s)[\vec \eta ]=0,\quad\quad
 \Rightarrow \vec \eta =0.
\label{IV10}
\end{eqnarray}

This explains why $m^{\mu}_{u(p_s)}(T_s,\vec \eta )$ is also called
the {\it mass dipole moment}.

Therefore, $x^{({\vec q}_{+})\mu}_s(T_s)$ is also simultaneously
the {\it Tulczyjew centroid}\cite{mul4,ehlers,mul8}
\footnote{Defined by $S_T^{\mu\nu}(T_s)[{\vec \eta }]
u_{\nu}(p_s)=0$, namely with $S^{or}_s=0$ in the momentum rest
frame.} and, due to ${\dot x}_s^{({\vec q}
_{+})\mu}(T_s)=u^{\mu}(p_s)$, also the {\it Pirani
centroid}\cite{pirani}
 \footnote{Defined by
$S_T^{\mu\nu}(T_s)[\vec \eta ] {\dot x}_{s\, \nu} =0$, namely with
$S^{or}_s=0$ in the instantaneous velocity rest frame.}. Usually,
in absence of a relation between 4-momentum and 4-velocity they
are different centroids \footnote{For instance, the so-called {\it
background Corinaldesi-Papapetrou centroid}\cite{cori} is defined
by the condition $S_T^{\mu\nu}(T_s)[\vec \eta ]v_{\nu}=0$, where
$v^{\mu}$ is a given fixed unit 4-vector.}.

Let us remark that   non-covariant centroids could also be connected
with the non-covariant external center of mass ${\tilde x}^{\mu}_s$
and the non-covariant external M\o ller center of energy.

\subsection{Quadrupoles.}

The {\it quadrupoles} correspond to $n=2$
\footnote{$t_T^{\mu_1\mu_2\mu\nu}=\epsilon^{\mu_1}_{r_1}(u(p_s))\epsilon^{\mu_2}_{r_2}(u(p_s))
\epsilon^{\mu}_A(u(p_s))\epsilon^{\nu}_B(u(p_s)) q_T^{r_1r_2AB}$,
$t_T^{\mu_1\mu_2\mu}{}_{\mu}=\epsilon^{\mu_1}_{r_1}(u(p_s))\epsilon^{\mu_2}_{r_2}(u(p_s))
\epsilon^{\mu}_A(u(p_s)) q_T^{r_1r_2A}{}_A$, ${\tilde t}_T^{\mu_1\mu_2}=
\epsilon^{\mu_1}_{r_1}(u(p_s))\epsilon^{\mu_2}_{r_2}(u(p_s)) q_T^{r_1r_2\tau\tau}$,
$p_T^{\mu_1\mu_2\mu}=\epsilon^{\mu_1}_{r_1}(u(p_s))\epsilon^{\mu_2}_{r_2}(u(p_s))
\epsilon^{\mu}_A(u(p_s))q_T^{r_1r_2A\tau}$.}:

\begin{eqnarray}
q_T^{r_1r_2AB}(T_s,\vec \eta )&=&\delta^A_{\tau}\delta^B_{\tau}
\sum_{i=1}^N[\eta_i^{r_1}(T_s)-\eta^{r_1}(T_s)][\eta_i^{r_2}(T_s)-\eta^{r_2}(T_s)]
\sqrt{m_i^2+{\vec \kappa}_i^2(T_s)}+\nonumber \\
&+&\delta^A_u\delta^B_v
\sum_{i=1}^N[\eta_i^{r_1}(T_s)-\eta^{r_1}(T_s)][\eta_i^{r_2}(T_s)-\eta^{r_2}(T_s)]
{{\kappa_i^u\kappa_i^v}\over {\sqrt{m_i^2+{\vec
\kappa}_i^2}}}(T_s)+\nonumber \\
&+&(\delta^A_{\tau}\delta^B_u+\delta^A_u\delta^B_{\tau})
\sum_{i=1}^N[\eta_i^{r_1}(T_s)-\eta^{r_1}(T_s)][\eta_i^{r_2}(T_s)-\eta^{r_2}(T_s)]
\kappa_i^u(T_s),\nonumber \\
 &&{}\nonumber \\
 q_T^{r_1r_2\tau\tau}(T_s,{\vec R}_{+})&=&\sum_{i=1}^N(\eta_i^{r_1}-R^{r_1}_{+})(\eta_i^{r_2}-
 R^{r_2}_{+}) H_i=\nonumber \\
 &=&{1\over N} \sum_{ijk}^{1..N}\sum_{ab}^{1..N-1} (\gamma_{ai}-\gamma_{aj})(\gamma_{bi}
 -\gamma_{bk}) \rho_a^{r_1}\rho_b^{r_2} {{H_iH_jH_k}\over {H_M^2}},\nonumber \\
 q_T^{r_1r_2u\tau}(T_s,{\vec R}_{+})&=&\sum_{i=1}^N (\eta_i^{r_1}-R^{r_1}_{+})
 (\eta_i^{r_2}-R^{r_2}_{+}) \kappa_i^u=\nonumber \\
 &=&{1\over N} \sum_{ijk}^{1..N}\sum_{ab}^{1..N-1}(\gamma_{ai}-\gamma_{aj})
 (\gamma_{bi}-\gamma_{bk}) \rho_a^{r_1}\rho_b^{r_2} {{\kappa_i^u H_jH_k}\over {H_M^2}},
 \nonumber \\
 q_T^{r_1r_2uv}(T_s,{\vec R}_{+})&=& \sum_{i=1}^N(\eta_i^{r_1}-R^{r_1}_{+})(\eta_i^{r_2}-
 R^{r_2}_{+}) {{\kappa_i^u\kappa_i^v}\over {H_i}}=\nonumber \\
 &=&{1\over N} \sum_{ijk}^{1..N}\sum_{ab}^{1..N-1}(\gamma_{ai}-\gamma_{aj})
 (\gamma_{bi}-\gamma_{bk}) \rho_a^{r_1}\rho_b^{r_2} {{\kappa_i^u\kappa_i^v H_jH_k}\over
 {H_i H_M^2}}.
\label{IV11}
\end{eqnarray}

Dixon's definition of {\it barycentric tensor of inertia} follows the
non-relativistic pattern   starting from the {\it mass quadrupole}

\begin{equation}
q_T^{r_1r_2\tau\tau}(T_s,\vec \eta )=
\sum_{i=1}^N[\eta_i^{r_1}\eta_i^{r_2}\sqrt{m_i^2+{\vec \kappa}_i^2}](T_s)
+M[\eta^{r_1}\eta^{r_2}-\eta^{r_1}R^{r_2}_{+}-\eta^{r_2}R^{r_1}_{+}](T_s),
\label{IV12}
\end{equation}

\begin{eqnarray}
I_{dixon}^{r_1r_2}(T_s)&=&\delta^{rs} \sum_u q_T^{uu\tau\tau}(T_s,0)-
q_T^{r_1r_2\tau\tau}(T_s,0)=\nonumber \\
&=&\sum_{i=1}^N[(\delta^{r_1r_2} {\vec
\eta}_i^2-\eta_i^{r_1}\eta_i^{r_2})\sqrt{m_i^2+{\vec \kappa}_i^2}](T_s)
\nonumber \\
&{\rightarrow}_{\alpha \rightarrow \infty}\,&   {1\over
N}\sum_{ijk}^{1..N}\sum_{ab}^{1..N-1}(\gamma_{ai}-\gamma_{aj})(\gamma_{bi}-\gamma_{bk})
\nonumber \\
&&[{\vec \rho}_{qa}\cdot {\vec
\rho}_{qb}\delta^{r_1r_2}-\rho_{qa}^{r_1}\rho_{qb}^{r_2}]{{H_i(\infty )H_j(\infty )H_k(\infty )}
\over {\Pi}}=\nonumber \\
&=&\sum_{ab}^{1..N-1}\Big( {1\over N}
\sum_{ijk}^{1..N}(\gamma_{ai}-\gamma_{aj})(\gamma_{bi}-\gamma_{bk})\nonumber \\
&&{{ \sqrt{m_i^2+N\sum_{de}\gamma_{di}\gamma_{ei}{\vec
\pi}_{qd}\cdot {\vec \pi}_{qe}}
 }\over {(\sum_{h=1}^N
\sqrt{m_h^2+N\sum_{de}\gamma_{dh}\gamma_{eh}{\vec \pi}_{qd}\cdot {\vec \pi}_{qe}})^2}}
 \nonumber \\
&&\sqrt{m_j^2+N\sum_{de}\gamma_{dj}\gamma_{ej}{\vec \pi}_{qd}\cdot
{\vec \pi}_{qe}}
\sqrt{m_k^2+N\sum_{de}\gamma_{dk}\gamma_{ek}{\vec \pi}_{qd}\cdot
{\vec \pi}_{qe}}   \Big) \nonumber \\
 &&[{\vec \rho}_{qa}\cdot {\vec \rho}_{qb}
\delta^{r_1r_2}-\rho_{qa}^{r_1}\rho_{qb}^{r_2}]\nonumber \\
&{\rightarrow}_{c \rightarrow \infty}\,&
\sum_{ab}^{1..N-1}\sum_{ijk}^{1..N}{{m_im_jm_k}\over {Nm^2}}
(\gamma_{ai}-\gamma_{aj})(\gamma_{bi}-\gamma_{bk})
{\vec \rho}_{qa}\cdot {\vec \rho}_{qb}
 \delta^{r_1r_2}-\rho_{qa}^{r_1}\rho_{qb}^{r_2}]\times \nonumber \\
 &&\times \Big[ 1+{1\over c} \Big( {{N\sum_{cd}^{1..N-1}\gamma_{ci}\gamma_{di}
 {\vec \pi}_{qc}\cdot {\vec \pi}_{qd}}\over {2m_i^2}}+{{N\sum_{cd}^{1..N-1}
 \gamma_{cj}\gamma_{dj}{\vec \pi}_{qc}\cdot {\vec \pi}_{qd}}\over {2m_j^2}}+
 \nonumber \\
 &+&{{N\sum_{cd}^{1..N-1}\gamma_{ck}\gamma_{dk}{\vec \pi}_{qc}\cdot
 {\vec \pi}_{qd}}\over {2m_k^2}} -{1\over m}\sum_{h=1}^N{{N\sum_{cd}^{1..N-1}
 \gamma_{ch}\gamma_{dh}{\vec \pi}_{qc}\cdot {\vec \pi}_{qd}}\over {m_h}}\Big) +O(1/c^2)\Big]
 =\nonumber \\
 &=&\sum_{ab}^{1..N-1} k_{ab} {\vec \rho}_{qa}\cdot {\vec \rho}_{qb}
 \delta^{r_1r_2}-\rho_{qa}^{r_1}\rho_{qb}^{r_2}] +O(1/c)=\nonumber \\
 &=& I^{r_1r_2}[{\vec q}_{nr}] + O(1/c).
\label{IV13}
\end{eqnarray}

In the non-relativistic limit we recover the  tensor of inertia of
Eqs.(\ref{a11}).

On the other hand, Thorne's definition of {\it barycentric tensor
of inertia}\cite{thorne} is

\begin{eqnarray}
I_{thorne}^{r_1r_2}(T_s)&=&\delta^{r_1r_2} \sum_u
q_T^{uuA}{}_A(T_s,0)- q_T^{r_1r_2A}{}_A(T_s,0)=\nonumber \\
 &=& \sum_{i=1}^N {{m_i^2 (\delta^{r_1r_2}{\vec \eta}_i^2-\eta_i^{r_1}\eta_i
 ^{r_2})}\over {\sqrt{m_i^2+{\vec \kappa}_i^2}}}(T_s)\nonumber \\
 &{\rightarrow}_{\alpha \rightarrow \infty}\,& {1\over N} \sum_{ijk}^{1..N}\sum_{ab}^{1..N-1}
 (\gamma_{ai}-\gamma_{aj})(\gamma_{bi}-\gamma_{bk})\nonumber \\
 && [{\vec \rho}_{qa}\cdot {\vec \rho}_{qb}
 \delta^{r_1r_2}-\rho_{qa}^{r_1}\rho_{qb}^{r_2}]{{m_i^2H_j(\infty )H_k(\infty )}\over
 {H_i(\infty ) \Pi}}=\nonumber \\
 &=& \sum_{ab}^{1..N-1} \Big( {c\over N} \sum_{ijk}^{1..N}(\gamma_{ai}-\gamma_{aj})
 (\gamma_{bi}-\gamma_{bk}) \nonumber \\
 &&{{ m_i^2
 \sqrt{m_j^2+N\sum_{de}\gamma_{dj}\gamma_{ej}{\vec \pi}_{qd}\cdot {\vec \pi}_{qe}}
 \sqrt{m_k^2+N\sum_{de}\gamma_{dk}\gamma_{ek}{\vec \pi}_{qd}\cdot {\vec \pi}_{qe}} }\over
 { \sqrt{m_i^2+N\sum_{de}\gamma_{di}\gamma_{ei}{\vec \pi}_{qd}\cdot {\vec \pi}_{qe}}
 (\sum_{h=1}^N\sqrt{m_h^2+N\sum_{de}\gamma_{dh}\gamma_{eh}
 {\vec \pi}_{qd}\cdot {\vec \pi}_{qe}})^2}} \Big) \nonumber \\
 &&[{\vec \rho}_{qa}\cdot {\vec \rho}_{qb}
 \delta^{r_1r_2}-\rho_{qa}^{r_1}\rho_{qb}^{r_2}]\nonumber \\
 &{\rightarrow}_{c \rightarrow \infty}\,&
\sum_{ab}^{1..N-1}\sum_{ijk}^{1..N}{{m_im_jm_k}\over {Nm^2}}
(\gamma_{ai}-\gamma_{aj})(\gamma_{bi}-\gamma_{bk})
{\vec \rho}_{qa}\cdot {\vec \rho}_{qb}
 \delta^{r_1r_2}-\rho_{qa}^{r_1}\rho_{qb}^{r_2}]\times \nonumber \\
 &&\times \Big[ 1+{1\over c} \Big( -{{N\sum_{cd}^{1..N-1}\gamma_{ci}\gamma_{di}
 {\vec \pi}_{qc}\cdot {\vec \pi}_{qd}}\over {2m_i^2}}+{{N\sum_{cd}^{1..N-1}
 \gamma_{cj}\gamma_{dj}{\vec \pi}_{qc}\cdot {\vec \pi}_{qd}}\over {2m_j^2}}+
 \nonumber \\
 &+&{{N\sum_{cd}^{1..N-1}\gamma_{ck}\gamma_{dk}{\vec \pi}_{qc}\cdot
 {\vec \pi}_{qd}}\over {2m_k^2}} -{1\over m}\sum_{h=1}^N{{N\sum_{cd}^{1..N-1}
 \gamma_{ch}\gamma_{dh}{\vec \pi}_{qc}\cdot {\vec \pi}_{qd}}\over {m_h}}\Big) +O(1/c^2)\Big]
 =\nonumber \\
 &=&\sum_{ab}^{1..N-1} k_{ab} {\vec \rho}_{qa}\cdot {\vec \rho}_{qb}
 \delta^{r_1r_2}-\rho_{qa}^{r_1}\rho_{qb}^{r_2}] +O(1/c)=\nonumber \\
 &=& I^{r_1r_2}[{\vec q}_{nr}] + O(1/c).
\label{IV14}
\end{eqnarray}

In this case too we recover the tensor of inertia of
Eq.(\ref{a11}). The Dixon and Thorne barycentric tensors of
inertia differ at the post-Newtonian level:\hfill\break
\hfill\break

\begin{eqnarray*}
I^{r_1r_2}_{dixon}(T_s)-I^{r_1r_2}_{thorne}(T_s)&=& {1\over c}
\sum_{ab}^{1..N-1}\sum_{ijk}^{1..N}{{m_jm_k}\over {Nm^2}}
(\gamma_{ai}-\gamma_{aj})(\gamma_{bi}-\gamma_{bk}) \nonumber \\
 &&\Big[{\vec \rho}_{qa}\cdot {\vec \rho}_{qb}
 \delta^{r_1r_2}-\rho_{qa}^{r_1}\rho_{qb}^{r_2}\Big] {{N\sum_{cd}^{1..N-1}
 \gamma_{ci}\gamma_{di}{\vec \pi}_{qc}\cdot {\vec \pi}_{qd}}\over {m_i}}+O(1/c^2).
 \end{eqnarray*}

\subsection{The Multipolar Expansion.}

In the next Section further types of Dixon's multipoles are
analyzed. It turns out that the {\it multipolar expansion}
(\ref{IV4}) can be rearranged with the help of the fact that
Hamilton equations imply $\partial_{\mu}T^{\mu\nu}\, {\buildrel
\circ \over =}\, 0$, so that from Eqs.(\ref{V8}) we get the {\it
multipolar expansion}

\begin{eqnarray}
&&T^{\mu\nu}[ x_s^{({\vec q}_{+})
\beta}(T_s)+\epsilon^{\beta}_r(u(p_s))\sigma^r ]=
T^{\mu\nu}[ w^{\beta}(T_s)+\epsilon^{\beta}_r(u(p_s))
(\sigma^r-\eta^r(T_s)) ]=\nonumber \\
 &&{}\nonumber \\
 &=& u^{(\mu}(p_s) \epsilon^{\nu )}_A(u(p_s)) [\delta^A_{\tau}
  M+\delta^A_u \kappa^u_{+}] \delta^3(\vec \sigma -\vec \eta
  (T_s))+\nonumber \\
  &+&{1\over 2} S_T^{\rho (\mu}(T_s)[\vec \eta ] u^{\nu )}(p_s)
   \epsilon^r_{\rho}(u(p_s)) {{\partial}\over {\partial \sigma^r}}
   \delta^3(\vec \sigma -\vec \eta (T_s))+\nonumber \\
   &+&\sum_{n=2}^{\infty} {{(-1)^n}\over {n!}} I_T^{\mu_1..\mu_n\mu\nu}(T_s,\vec \eta )
   \epsilon^{r_1}_{\mu_1}(u(p_s))..\epsilon^{r_n}_{\mu_n}(u(p_s))
   {{\partial^n}\over {\partial \sigma^{r_1}..\partial \sigma^{r_n}}}
   \delta^3(\vec \sigma -\vec \eta (T_s)),
\label{IV15}
\end{eqnarray}

\noindent where for $n\geq 2$ and $\vec \eta =0$
$\quad I_T^{\mu_1..\mu_n\mu\nu}(T_s)={{4(n-1)}\over {n+1}}
J_T^{(\mu_1..\mu_{n-1} | \mu | \mu_n)\nu}(T_s)$, with
$J_T^{\mu_1..\mu_n\mu\nu\rho\sigma}(T_s)$ being the Dixon {\it
$2^{2+n}$-pole inertial moment tensors} given in Eqs.(\ref{V10}).

The equations $\partial_{\mu}T^{\mu\nu}\, {\buildrel \circ \over
=}\, 0$ imply the {\it Papapetrou-Dixon-Souriau equations of
motion} \cite{mul5,dixon1,souriau,mul14} for the total momentum
$P^{\mu}_T(T_s)=\epsilon^{\mu}_A(u(p_s)) q_T^{A\tau}(T_s)\approx
p^{\mu}_s$ and the spin tensor $S^{\mu\nu}_T(T_s)[\vec \eta =0]$
restricted to positive energy particles   [see Eqs.(\ref{V4}) and
(\ref{V7})]

\begin{eqnarray}
{{d P^{\mu}_T(T_s)}\over {dT_s}}\, &{\buildrel \circ \over =}\,&
0,\nonumber \\
 {{d S^{\mu\nu}_T(T_s)[\vec \eta =0]}\over {dT_s}}\, &{\buildrel
 \circ \over =}\,& 2 P^{[\mu}_T(T_s) u^{\nu ]}(p_s)=2 \kappa^u_{+} \epsilon^{[\mu}_u(u(p_s))
 u^{\nu ]}(p_s) \approx 0.
\label{IV16}
\end{eqnarray}

\vfill\eject

\section{More on Dixon's Multipoles.}

In this Section we shall consider multipoles with respect to the
origin, i.e. with $\vec \eta =0$ [we use the notation
$t_T^{\mu_1...\mu_n\mu\nu}(T_s, 0)= t_T^{\mu_1...\mu_n\mu\nu}(T_s)$]

As shown in Ref.\cite{dixon}, if a field has a compact support W on
the Wigner hyperplanes $\Sigma_{W\tau}$ and if $f(x)$ is a
$C^{\infty}$ complex-valued scalar function on Minkowski spacetime
with compact support \footnote{So that its Fourier transform $\tilde
f(k)=\int d^4x f(x) e^{ik\cdot x}$ is a slowly increasing entire
analytic function on Minkowski spacetime ($|(x^o+iy^o)
^{q_o}...(x^3+iy^3)^{q_3}f(x^{\mu}+iy^{\mu})| < C_{q_o...q_3} e^{a_o|y^o|+...+
a_3|y^3|}$, $a_{\mu} > 0$, $q_{\mu}$ positive integers for every $\mu$ and
$C_{q_o...q_3} > 0$), whose inverse is $f(x)=\int {{d^4k}\over {(2\pi )^4}}
\tilde f(k)e^{-ik\cdot x}$.}, we have

\begin{eqnarray}
< T^{\mu\nu},f >&=& \int d^4x T^{\mu\nu}(x) f(x)=\nonumber \\
&=&\int dT_s\int d^3\sigma f(x_s+\delta x_s)T^{\mu\nu}[x_s(T_s)+\delta x_s(\vec
\sigma )][\phi ]=\nonumber \\
&=&\int dT_s \int d^3\sigma \int {{d^4k}\over {(2\pi )^4}} \tilde f(k)
e^{-ik\cdot [x_s(T_s)+\delta x_s(\vec \sigma )]}T^{\mu\nu}
[x_s(T_s)+\delta x_s(\vec \sigma )][\phi ]=\nonumber \\
&=&\int dT_s\int {{d^4k}\over {(2\pi )^4}} \tilde f(k) e^{-ik\cdot x_s(T_s)}
\int d^3\sigma T^{\mu\nu}[x_s(T_s)+\delta x_s(\vec \sigma )][\phi ]\nonumber \\
&&\sum_{n=0}^{\infty} {{(-i)^n}\over {n!}} [k_{\mu}\epsilon^{\mu}_u(u(p_s))
\sigma^u]^n=\nonumber \\
&=&\int T_s\int {{d^4k}\over {(2\pi )^4}} \tilde f(k) e^{-ik\cdot x_s(T_s)}
\sum_{n=0}^{\infty} {{(-i)^n}\over {n!}} k_{\mu_1}...k_{\mu_n} t_T^{\mu_1...
\mu_n\mu\nu}(T_s),
\label{V1}
\end{eqnarray}

\noindent and, but only for $f(x)$ analytic in W \cite{dixon} {\footnote{See this
paper for related results of Mathisson and
Tulczyjev\cite{mul1,mul2,mul3,mul4}.}, we get

\begin{eqnarray}
< T^{\mu\nu},f >&=& \int dT_s \sum_{n=0}^{\infty}{1\over {n!}} t_T^{\mu_1...
\mu_n\mu\nu}(T_s) {{\partial^nf(x)}\over {\partial x^{\mu_1}...\partial x^{\mu
_n}}}{|}_{x=x_s(T_s)},\nonumber \\
&&\Downarrow \nonumber \\
T^{\mu\nu}(x)&=& \sum_{n=0}^{\infty}{{(-1)^n}\over {n!}}
{{\partial^n}\over {\partial x^{\mu_1}...\partial x^{\mu_n}}} \int dT_s
\delta^4(x-x_s(T_s)) t_T^{\mu_1...\mu_n\mu\nu}(T_s).
\label{V2}
\end{eqnarray}

For a N particle system this equation may be rewritten as
Eq.(\ref{IV4}).

For non-analytic functions $f(x)$  we have

\begin{eqnarray}
< T^{\mu\nu},f > &=& \int dT_s \sum^N_{n=0} {1\over {n!}} t_T^{\mu_1...\mu_n
\mu\nu}(T_s) {{\partial^nf(x)}\over {\partial x^{\mu_1}...\partial x^{\mu
_n}}}{|}_{x=x_s(T_s)}+\nonumber \\
&+&\int dT_s \int {{d^4k}\over {(2\pi )^4}} \tilde f(k) e^{-ik\cdot x_s(T_s)}
\sum_{n=N+1}^{\infty} {{(-i)^n}\over {n!}} k_{\mu_1}...k_{\mu_n} t_T^{\mu_1...
\mu_n\mu\nu}(T_s),
\label{V3}
\end{eqnarray}

\noindent and, as shown in Ref.\cite{dixon}, from the knowledge of
the moments $t_T^{\mu_1...\mu_n\mu}(T_s)$ for all $n > N$ we can
get $T^{\mu\nu}(x)$ and, therefore, all the moments with $n\leq
N$.

The Hamilton equations  imply \footnote{In Ref.\cite{dixon} this is a
consequence of $\partial_{\mu}T^{\mu\nu}\, {\buildrel \circ \over =}\,
0$.} imply (we omit the dependence on $\vec \eta =0$ of the
multipoles)

\begin{eqnarray}
{{dp_T^{\mu}(T_s)}\over {dT_s}}\, &{\buildrel \circ \over =}\,& 0,\quad
for\, n=0,\nonumber \\
{{d p_T^{\mu_1...\mu_n\mu}(T_s)}\over {dT_s}}\, &{\buildrel \circ \over =}\,&
-nu^{(\mu_1}(p_s) p_T^{\mu_2...\mu_n)\mu}(T_s)+n t_T^{(\mu_1...\mu_n)\mu}(T_s),
\quad n\geq 1.
\label{V4}
\end{eqnarray}

If we define for $n \geq 1$ [$S^{\mu\nu}_T=2 p_T^{[\mu\nu ]}=2
c_T^{\mu\nu}$]

\begin{eqnarray}
b_T^{\mu_1...\mu_n\mu}(T_s)&=&p_T^{(\mu_1...\mu_n\mu )}(T_s)=\nonumber \\
&=&
\epsilon^{(\mu_1}_{r_1}(u(p_s)).... \epsilon^{\mu_n}_{r_n}(u(p_s))\epsilon^{\mu )}_A(u(p_s))
q_T^{r_1..r_nA\tau}(T_s) ,\nonumber \\
 &&{}\nonumber \\
 \epsilon^{r_1}_{\mu_1}(u(p_s))....\epsilon^{r_n}_{\mu_n}(u(p_s)) b_T^{\mu_1...\mu_n\mu}(T_s)
 &=&{1\over {n+1}} u^{\mu}(p_s) q_T^{r_1...r_n\tau\tau}(T_s)+\epsilon^{\mu}_r(u(p_s))
 q_T^{(r_1...r_nr)\tau}(T_s),\nonumber \\
 &&{}\nonumber \\
c_T^{\mu_1...\mu_n\mu}(T_s)&=&c_T^{(\mu_1...\mu_n
)\mu}(T_s)=p_T^{\mu_1...
\mu_n\mu}(T_s)-p_T^{(\mu_1...\mu_n\mu )}(T_s)=\nonumber \\
&=&[\epsilon^{\mu_1}_{r_1}(u(p_s))...\epsilon^{\mu_n}_{r_n}\epsilon^{\mu}_A(u(p_s))-
    \nonumber \\
 &-&\epsilon^{(\mu_1}_{r_1}(u(p_s))...\epsilon^{\mu_n}_{r_n}(u(p_s))
\epsilon_A^{\mu )}(u(p_s))] q_T^{r_1..r_nA\tau}(T_s),\nonumber \\
 &&{}\nonumber \\
&&c_T^{(\mu_1...\mu_n\mu )}(T_s)=0,\nonumber \\
 \epsilon^{r_1}_{\mu_1}(u(p_s))....\epsilon^{r_n}_{\mu_n}(u(p_s)) c_T^{\mu_1...\mu_n\mu}(T_s)
 &=& {n\over {n+1}}u^{\mu}(p_s) q_T^{r_1...r_n\tau\tau}(T_s) +\nonumber \\
 &+&\epsilon^{\mu}_r(u(p_s)) [ q_T^{r_1...r_nr\tau}(T_s) - q_T^{(r_1...r_nr)\tau}(T_s)],
\label{V5}
\end{eqnarray}

\noindent and then for $n\geq 2$

\begin{eqnarray}
d_T^{\mu_1...\mu_n\mu\nu}(T_s)&=&d_T^{(\mu_1...\mu_n)(\mu\nu )}(T_s)=
t_T^{\mu_1...\mu_n\mu\nu}(T_s)-\nonumber \\
 &-&{{n+1}\over
n}[t_T^{(\mu_1...\mu_n\mu )\nu}(T_s)+t_T^{(\mu_1...\mu_n\nu )\mu}
(T_s)]+\nonumber \\
 &+&{{n+2}\over n}t_T^{(\mu_1...\mu_n\mu\nu
)}(T_s)=\nonumber
\\ &=&\Big[ \epsilon^{\mu_1}_{r_1} ... \epsilon^{\mu_n}_{r_n}
\epsilon^{\mu}_A \epsilon^{\nu}_B-
 {{n+1}\over n}\Big( \epsilon^{(\mu_1}_{r_1} ... \epsilon^{\mu_n}_{r_n} \epsilon^{\mu )}_A
 \epsilon^{\nu}_B+\nonumber \\
 &+&\epsilon^{(\mu_1}_{r_1} ... \epsilon^{\mu_n}_{r_n} \epsilon^{\nu )}_B
 \epsilon^{\mu}_A\Big) +{{n+2}\over n} \epsilon^{(\mu_1}_{r_1} .. \epsilon^{\mu_n}_{r_n}
 \epsilon^{\mu}_A \epsilon^{\nu )}_B\Big] (u(p_s))\nonumber \\
 && q_T^{r_1..r_nAB}(T_s),\nonumber \\
 &&{}\nonumber \\
 &&d_T^{(\mu_1...\mu_n\mu )\nu}(T_s)=0,\nonumber \\
 &&{}\nonumber \\
 \epsilon^{r_1}_{\mu_1}(u(p_s))....\epsilon^{r_n}_{\mu_n}(u(p_s)) d_T^{\mu_1...\mu_n\mu\nu}(T_s)
 &=& {{n-1}\over {n+1}} u^{\mu}(p_s)u^{\nu}(p_s) q_T^{r_1...r_n\tau\tau}(T_s)+\nonumber \\
 &+&{1\over n} [u^{\mu}(p_s)\epsilon^{\nu}_r(u(p_s))+u^{\nu}(p_s)\epsilon^{\mu}_r(u(p_s))]
   \nonumber \\
 &&[(n-1) q_T^{r_1...r_nr\tau}(T_s)+ q_T^{(r_1...r_nr)\tau}(T_s)]+\nonumber \\
 &+&\epsilon^{\mu}_{s_1}(u(p_s))\epsilon^{\nu}_{s_2}(u(p_s)) [q_T^{r_1...r_ns_1s_2}(T_s)
 -\nonumber \\
  &-&{{n+1}\over n}( q_T^{(r_1...r_ns_1)s_2}(T_s)+ q_T^{(r_1...r_ns_2)s_1}(T_s)) +\nonumber \\
 &+& q_T^{(r_1...r_ns_1s_2)}(T_s)].
\label{V6}
\end{eqnarray}

 Eqs.(\ref{V4}) may be rewritten in the form

\begin{eqnarray}
&&1)\quad n=1\nonumber \\ &&{}\nonumber \\
t^{\mu\nu}_T(T_s)&=&t^{(\mu\nu )}_T(T_s)\, {\buildrel \circ \over =}\,
p_T^{\mu}(T_s)u^{\nu }(p_s)+{1\over 2}{d\over
{dT_s}}(S^{\mu\nu}_T(T_s)+2b_T
^{\mu\nu}(T_s)),\nonumber \\
&&\Downarrow \nonumber \\
 t^{\mu\nu}_T(T_s)\, &{\buildrel \circ \over
=}\,&p_T^{(\mu}(T_s)u^{\nu )}(p_s) +{d\over
{dT_s}}b_T^{\mu\nu}(T_s)=M u^{\mu}(p_s)u^{\nu}(p_s)+\nonumber \\
 &+&\kappa^r_{+} [u^{(\mu}(p_s)\epsilon^{\nu )}_r(u(p_s))+
\epsilon^{(\mu}_r(u(p_s))u^{\nu )}(p_s)]+ \nonumber \\
 &&\epsilon^{(\mu}_r(u(p_s))\epsilon^{\nu )}_s(u(p_s))\sum_{i=1}^N
 {{\kappa^u_i \kappa^v_i}\over {\sqrt{m_i^2+{\vec \kappa}_i^2}}},\nonumber \\
 {d\over {dT_s}}S^{\mu\nu}_T(T_s)\, &{\buildrel \circ \over
=}\,&2p_T^{[\mu}(T_s) u^{\nu
]}(p_s)=2\kappa^r_{+}\epsilon^{[\mu}_r(u(p_s))u^{\nu ]}(p_s) \approx
0,\nonumber
\\ &&{}\nonumber \\
 &&2)\quad n=2\quad [identity\,\,
t_T^{\rho\mu\nu}=t_T^{(\rho\mu )\nu}+t_T^{(\rho
\nu )\mu}+t_T^{(\mu\nu )\rho}]\nonumber \\
&&{}\nonumber \\
2t_T^{(\rho\mu )\nu}(T_s)\, &{\buildrel \circ \over =}\,& 2u^{(\rho}(p_s)b_T
^{\mu )\nu}(T_s)+u^{(\rho}(p_s)S_T^{\mu )\nu}(T_s)+{d\over {dT_s}}(b_T^{\rho\mu\nu}
(T_s)+c_T^{\rho\mu\nu}(T_s)),\nonumber \\
&&\Downarrow \nonumber \\
t_T^{\rho\mu\nu}(T_s)\, &{\buildrel \circ \over =}\,&u^{\rho}(p_s)b_T
^{\mu\nu}(T_s)+S_T^{\rho (\mu}(T_s)u^{\nu )}(p_s)+{d\over {dT_s}}({1\over 2}b_T
^{\rho\mu\nu}(T_s)-c_T^{\rho\mu\nu}(T_s)),\nonumber \\
&&{}\nonumber \\
&&3) \quad n \geq 3 \nonumber \\
&&{}\nonumber \\
t_T^{\mu_1...\mu_n\mu\nu}(T_s)\, &{\buildrel \circ \over =}\,& d_T^{\mu_1...
\mu_n\mu\nu}(T_s)+u^{(\mu_1}(p_s)b_T^{\mu_2...\mu_n)\mu\nu}(T_s)+2u^{(\mu
_1}(p_s)c_T^{\mu_2...\mu_n)(\mu\nu )}(T_s)+\nonumber \\
&=&{2\over n}c_T^{\mu_1...\mu_n(\mu}(T_s)u^{\nu )}(p_s)+{d\over {dT_s}}
[{1\over {n+1}}b_T^{\mu_1...\mu_n\mu\nu}(T_s)+{2\over n}c_T^{\mu_1...\mu
_n(\mu\nu )}(T_s)],
\label{V7}
\end{eqnarray}

 This allows  to rewrite $< T^{\mu\nu},f >$ in the following form\cite{dixon}

\begin{eqnarray}
< T^{\mu\nu},f > &=&\int dT_s \int {{d^4k}\over {(2\pi )^4}} \tilde
f(k) e^{-ik\cdot x_s(T_s)} \Big[ u^{(\mu}(p_s)p_T^{\nu
)}(T_s)-ik_{\rho}S^{\rho (\mu}_T(T_s) u^{\nu )}(p_s)+\nonumber \\
&+&\sum_{n=2}^{\infty}{{(-i)^n}\over {n!}} k_{\rho_1}...k_{\rho_n}
I_T^{\rho_1
...\rho_n\mu\nu}(T_s)\Big],
\label{V8}
\end{eqnarray}

\noindent with

\begin{eqnarray}
I_T^{\mu_1...\mu_n\mu\nu}(T_s)&=&I_T^{(\mu_1...\mu_n )(\mu\nu )}(T_s)=
d_T^{\mu_1...\mu_n\mu\nu}(T_s)-\nonumber \\
 &-&{2\over {n-1}}u^{(\mu_1}(p_s)c_T^{\mu_2...\mu_n )(\mu\nu )}(T_s)+\nonumber \\
 &+&{2\over n}
c_T^{\mu_1...\mu_n(\mu}(T_s)u^{\nu )}(p_s)=\nonumber \\
  &=&\Big[
\epsilon^{\mu_1}_{r_1}...\epsilon^{\mu_n}_{r_n} \epsilon^{\mu}_A
\epsilon^{\nu}_B-
 {{n+1}\over n}\Big( \epsilon^{(\mu_1}_{r_1}...\epsilon^{\mu_n}_{r_n} \epsilon^{\mu )}_A
 \epsilon^{\nu}_B+\nonumber \\
 &+&\epsilon^{(\mu_1}_{r_1}...\epsilon^{\mu_n}_{r_n} \epsilon^{\nu )}_B \epsilon^{\mu}_A\Big) +
 {{n+2}\over n} \epsilon^{(\mu_1}_{r_1}...\epsilon^{\mu_n}_{r_n}
 \epsilon^{\mu}_A \epsilon^{\nu )}_B\Big] (u(p_s))\nonumber \\
 && q_T^{r_1..r_nAB}(T_s)-\nonumber \\
 &-&\Big[ {2\over {n-1}} u^{(\mu_1}(p_s) \Big( \epsilon^{\mu_2}_{r_1}...
 \epsilon^{\mu_n)}_{r_{n-1}} \epsilon^{(\mu}_{r_n} \epsilon^{\nu )}_A-
 \epsilon^{(\mu_2}_{r_1}...\epsilon^{\mu_n)}_{r_{n-1}} \epsilon^{(\mu}_{r_n}
 \epsilon^{\nu ))}_A\Big)-\nonumber \\
 &-&{2\over n} \Big( \epsilon^{\mu_1}_{r_1}...\epsilon^{\mu_n}_{r_n} \epsilon^{(\mu}_A-
 \epsilon^{(\mu_1}_{r_1}... \epsilon^{\mu_n}_{r_n} \epsilon^{(\mu )}_A u^{\nu )}(p_s)
 \Big] (u(p_s))\nonumber \\
 && q_T^{r_1..r_nA\tau}(T_s),\nonumber \\
 &&{}\nonumber \\
 &&I_T^{(\mu_1...\mu_n\mu )\nu}(T_s)=0,\nonumber \\
 &&{}\nonumber \\
 \epsilon^{r_1}_{\mu_1}(u(p_s))....\epsilon^{r_n}_{\mu_n}(u(p_s)) I_T^{\mu_1...\mu_n\mu\nu}(T_s)
 &=& {{n+3}\over {n+1}} u^{\mu}(p_s)u^{\nu}(p_s) q_T^{r_1...r_n\tau\tau}(T_s)+\nonumber \\
 &+&{1\over n}[u^{\mu}(p_s)\epsilon^{\nu}_r(u(p_s))+u^{\nu}(p_s)\epsilon^{\mu}_r(u(p_s))]
 q_T^{r_1...r_nr\tau}(T_s)+\nonumber \\
 &+&\epsilon^{\mu}_{s_1}(u(p_s))\epsilon^{\nu}_{s_2}(u(p_s)) [q_T^{r_1...r_ns_1s_2}(T_s)
 -\nonumber \\
 &-&{{n+1}\over n}( q_T^{(r_1...r_ns_1)s_2}(T_s)+ q_T^{(r_1...r_ns_2)s_1}(T_s)) +
    \nonumber \\
 &+&q_T^{(r_1...r_ns_1s_2)}(T_s)].
\label{V9}
\end{eqnarray}

For a N particle system Eq.(\ref{V8}) implies Eq.(\ref{IV15}).

Finally, a set of multipoles equivalent to the
$I_T^{\mu_1..\mu_n\mu\nu}$ is \footnote{$A^{[\mu [\rho \nu ]\sigma
]}\, {\buildrel {def} \over =}\, {1\over 4}
(A^{\mu\nu\rho\sigma}-A^{\nu\rho\mu\sigma}-A^{\mu\sigma\nu\rho}+A^{\nu\sigma\mu\rho})$.}

\begin{eqnarray}
&&for\quad n \geq 0\nonumber \\ &&{}\nonumber \\
J_T^{\mu_1...\mu_n\mu\nu\rho\sigma}(T_s)&=&J_T^{(\mu_1...\mu_n)[\mu\nu
][\rho \sigma ]}(T_s)= I_T^{\mu_1...\mu_n[\mu [\rho\nu ]\sigma
]}(T_s)=\nonumber \\ &=&t_T^{\mu_1...\mu_n[\mu [\rho\nu ]\sigma
]}(T_s)-\nonumber \\
 &-&{1\over {n+1}}\Big[ u^{[\mu}(p_s)p_T^{\nu ]\mu_1...\mu_n[\rho\sigma
]}(T_s)+\nonumber \\
 &+&u^{[\rho}(p_s)p_T^{\sigma ]\mu_1...\mu_n[\mu\nu ]}(T_s)\Big]=\nonumber\\
&=&\Big[ \epsilon^{\mu_1}_{r_1} .. \epsilon^{\mu_n}_{r_n} \epsilon^{[\mu}_r
 \epsilon^{[\rho}_s \epsilon^{\nu ]}_A \epsilon^{\sigma ]}_B\Big] (u(p_s))
 q_T^{r_1..r_nAB}(T_s)-\nonumber \\
  &-&{1\over {n+1}}\Big[ u^{[\mu}(p_s) \epsilon^{\nu ]}_r(u(p_s)) \epsilon^{[\rho}_s(u(p_s))
  \epsilon^{\sigma ]}_A(u(p_s))+\nonumber \\
   &+&u^{[\rho}(p_s) \epsilon^{\sigma ]}_r(u(p_s))
  \epsilon^{[\mu}_s(u(p_s)) \epsilon^{\nu ]}_A(u(p_s))\Big]\nonumber \\
  &&\epsilon^{\mu_1}_{r_1}(u(p_s))...\epsilon^{\mu_n}_{r_n}(u(p_s))
  q_T^{rr_1..r_nsA\tau}(T_s),\nonumber \\
 &&{}\nonumber \\
 (n+4)(3n+5) && linearly\, independent\, components,\nonumber \\
&&{}\nonumber \\
u_{\mu_1}(p_s)J_T^{\mu_1...\mu_n\mu\nu\rho\sigma}(T_s)&=&
J_T^{\mu_1...\mu_{n-1}(\mu_n\mu\nu )\rho\sigma}(T_s)=0,\quad\quad
for\, n \geq 1,\nonumber \\
 &&{}\nonumber \\
 I_T^{\mu_1...\mu_n\mu\nu}(T_s)&=&{{4(n-1)}\over {n+1}}
J_T^{(\mu_1...\mu_{n-1}|
\mu |\mu_n)\nu}(T_s),\quad\quad for\, n \geq 2,\nonumber \\
 &&{}\nonumber \\
  \epsilon^{r_1}_{\mu_1}(u(p_s))....\epsilon^{r_n}_{\mu_n}(u(p_s))
  J_T^{\mu_1...\mu_n\mu\nu\rho\sigma}(T_s) &=& \Big[ \epsilon^{[\mu}_r
 \epsilon^{[\rho}_s \epsilon^{\nu ]}_A \epsilon^{\sigma ]}_B\Big] (u(p_s))
 q_T^{r_1..r_nAB}(T_s)-\nonumber \\
  &-&{1\over {n+1}}\Big[ u^{[\mu}(p_s) \epsilon^{\nu ]}_r(u(p_s)) \epsilon^{[\rho}_s(u(p_s))
  \epsilon^{\sigma ]}_A(u(p_s))+\nonumber \\
   &+&u^{[\rho}(p_s) \epsilon^{\sigma ]}_r(u(p_s))
  \epsilon^{[\mu}_s(u(p_s)) \epsilon^{\nu ]}_A(u(p_s))\Big]
  q_T^{rr_1..r_nsA\tau}(T_s).\nonumber \\
  &&{}
\label{V10}
\end{eqnarray}

The $J_T^{\mu_1..\mu_n\mu\nu\rho\sigma}$ are the Dixon {\it
$2^{n+2}$-pole inertial moment tensors} of the extended system:
they (or equivalently the $I_T^{\mu_1...\mu_n\mu\nu}$'s) determine
its energy-momentum tensor together with the {\it monopole}
$P^{\mu}_T$ and the {\it spin dipole} $S^{\mu\nu}_T$. The
equations $\partial_{\mu} T^{\mu\nu}\, {\buildrel \circ \over =}\,
0$ are satisfied due to the equations of motion (\ref{V7}) for
$P^{\mu}_T$ and $S^{\mu\nu}_T$ \footnote{The so called {\it
Papapetrou-Dixon-Souriau equations} given in Eq.(\ref{IV16}).}
without any need of the equations of motion for the
$J_T^{\mu_1..\mu_n\mu\nu\rho\sigma}$. When all the multipoles
$J_T^{\mu_1..\mu_n\mu\nu\rho\sigma}$ are zero (or negligible) one
speaks of a {\it pole-dipole} system.

On the Wigner hyperplane the content  of these {\it $2^{n+2}$-pole
inertial moment tensors} is replaced by the {\it Euclidean Cartesian
tensors} $q_T^{r_1...r_n\tau\tau}$, $q_T^{r_1...r_nr\tau}$,
$q_T^{r_1...r_nrs}$. As shown in Appendix B we can decompose these
Cartesian tensors in their irreducible STF (symmetric trace-free)
parts (the {\it STF tensors}).

The {\it multipolar expansion} (\ref{IV15}) may be rewritten as

\bea
&&T^{\mu\nu}[ x_s^{({\vec q}_{+})
\beta}(T_s)+\epsilon^{\beta}_r(u(p_s))\sigma^r ]=
T^{\mu\nu}[ w^{\beta}(T_s)+\epsilon^{\beta}_r(u(p_s))
(\sigma^r-\eta^r(T_s))]=\nonumber \\
 &&{}\nonumber \\
 &=& u^{(\mu}(p_s) \epsilon^{\nu )}_A(u(p_s)) [\delta^A_{\tau}
  M+\delta^A_u \kappa^u_{+}] \delta^3(\vec \sigma -\vec \eta
  (T_s))+\nonumber \\
  &+&{1\over 2} S_T^{\rho (\mu}(T_s)[\vec \eta ] u^{\nu )}(p_s)
   \epsilon^r_{\rho}(u(p_s)) {{\partial}\over {\partial \sigma^r}}
   \delta^3(\vec \sigma -\vec \eta (T_s))+\nonumber \\
   &+&\sum_{n=2}^{\infty} {{(-1)^n}\over {n!}} \Big[
 {{n+3}\over {n+1}} u^{\mu}(p_s)u^{\nu}(p_s) q_T^{r_1...r_n\tau\tau}(T_s,\vec \eta )+\nonumber \\
 &+&{1\over n}[u^{\mu}(p_s)\epsilon^{\nu}_r(u(p_s))+u^{\nu}(p_s)\epsilon^{\mu}_r(u(p_s))]
 q_T^{r_1...r_nr\tau}(T_s,\vec \eta )+\nonumber \\
 &+&\epsilon^{\mu}_{s_1}(u(p_s))\epsilon^{\nu}_{s_2}(u(p_s))
 [q_T^{r_1...r_ns_1s_2}(T_s,\vec \eta )-\nonumber \\
 &-&{{n+1}\over n}( q_T^{(r_1...r_ns_1)s_2}(T_s,\vec \eta )+
 q_T^{(r_1...r_ns_2)s_1}(T_s,\vec \eta )) +
 q_T^{(r_1...r_ns_1s_2)}(T_s, \vec \eta )]\Big] \nonumber \\
 &&{{\partial^n}\over {\partial \sigma^{r_1}..\partial \sigma^{r_n}}}
   \delta^3(\vec \sigma -\vec \eta (T_s)).
\label{V11}
\eea

\vfill\eject

\section{Conclusions.}

In this paper we have completed our study of the relativistic
kinematics of the system of N scalar positive-energy particles in
the rest-frame instant form of dynamics on Wigner hyperplanes
orthogonal to the system total 4-momentum initiated in
Ref.\cite{iten1}.

We have evaluated the energy momentum tensor of the system on the
Wigner hyperplane and then  determined Dixon's multipoles for the
N-body problem with respect to the {\it internal} 3-center of mass
located at the origin of the Wigner hyperplane \footnote{So that
the origin is also the Fokker-Pryce center of inertia and Pirani
and Tulczyjew centroids of the system.}. In the rest-frame instant
form of dynamics these multipoles are {\it Cartesian
(Wigner-covariant) Euclidean tensors}. While the study of the {\it
monopole} and {\it dipole} moments in the rest frame gives
informations on the mass, the spin and the {\it internal} 3-center
of mass, the {\it quadrupole} moment give the only (even if not
unique) way to introduce the concept of {\it barycentric tensor of
inertia} for extended systems in special relativity.

Finally, let us observe that, by exploiting the canonical spin
bases of Refs.\cite{iten2,iten1}, after the elimination of the
{internal} 3-center of mass (${\vec q}_{+}={\vec \kappa}_{+}=0$)
the Cartesian multipoles $q_T^{r_1...r_nAB}$ can be expressed in
terms of 6 {\it orientational} variables (the {\it spin vector}
and the three {\it Euler angles} identifying the dynamical body
frame) and of $6N-6$ ({\it rotational scalar}) {\it shape}
variables.

\vfill\eject

\appendix

\section{Non-Relativistic Multipolar Expansions for N Free  Particles.}

In the review paper in Ref.\cite{dixon1} there is a study of the
Newtonian multipolar expansions for a continuum isentropic
distribution of matter characterized by a mass density $\rho
(t,\vec \sigma )$,  a velocity field $U^r(t,\vec \sigma )$
\footnote{$\rho (t,\vec \sigma ) \vec U(t,\vec \sigma )$ is the
momentum density.} and  a stress tensor $\sigma^{rs}(t,\vec \sigma
)$. If the system is isolated, the only dynanical equations are
the mass conservation and the continuum equations of motion
respectively

\bea
&&{{\partial \rho (t,\vec \sigma )}\over {\partial t}}-{{\partial \rho
(t,\vec \sigma ) U^r(t,\vec \sigma )}\over {\partial
\sigma^r}}=0,\nonumber  \\
 &&{{\partial \rho (t,\vec \sigma )U^r(t,\vec \sigma )}\over {\partial
t}}-{{\partial [\rho U^r U^s -\sigma^{rs}](t,\vec \sigma )}\over
{\partial \sigma^s}}\, {\buildrel \circ \over =}\, 0.
\label{a1}
\eea

We can adapt this description to an isolated system of N particles in
the following way. The mass density

\begin{equation}
\rho (t,\vec \sigma )=\sum_{i=1}^N m_i \delta^3(\vec \sigma -{\vec \eta}_i(t)),
\label{a2}
\end{equation}

\noindent satisfies

\begin{equation}
{{\partial \rho (t,\vec \sigma )}\over {\partial t}}=-\sum_{i=1}^Nm_i
{\dot {\vec \eta}}_i(t)\cdot {\vec \partial}_{{\vec \eta}_i}
\delta^3(\vec \sigma -{\vec \eta}_i(t))\, {\buildrel {def} \over =}\,
{{\partial}\over {\partial \sigma^r}}[\rho U^r](t,\vec \sigma ),
\label{a3}
\end{equation}

\noindent while the momentum density \footnote{This can be taken as the
definitory equation for the velocity field, even if strictly speaking
we do not need it in what follows.}

\begin{equation}
\rho (t,\vec \sigma ) U^r(t,\vec \sigma )= \sum_{i=1}^N m_i {\dot
{\vec \eta}}_i(t) \delta^3(\vec \sigma -{\vec \eta}_i(t )),
\label{a4}
\end{equation}

The associated constant of motion is the total mass $m=\sum_{i=1}^N$.

If we define a function $\zeta (\vec \sigma ,{\vec \eta}_i)$
concentrated in the N points ${\vec \eta}_i$, i=1,..,N, such that
$\zeta (\vec
\sigma ,{\vec \eta}_i)=0$ for $\vec \sigma \not= {\vec
\eta}_i$ and $\zeta ({\vec \eta}_i,{\vec \eta}_j)=\delta_{ij}$
\footnote{It is a limiting concept deriving from the characteristic function of a
manifold.}, then the velocity field associated to N particles becomes

\begin{equation}
\vec U(t,\vec \sigma )= \sum_{i=1}^N {\dot {\vec \eta}}_i(t)
\zeta (\vec \sigma ,{\vec \eta}_i(t)).
\label{a5}
\end{equation}

The continuum equations of motion are replaced by

\begin{eqnarray}
&&{{\partial}\over {\partial t}} [\rho (t,\vec \sigma ) U^r(t,\vec
\sigma )]\, {\buildrel \circ \over =}\,
{{\partial}\over {\partial
\sigma^s}} \sum_{i=1}^N m_i {\dot
\eta}^r_i(t) {\dot \eta}^s_i(t) \delta^3(\vec \sigma -{\vec \eta}_i(t))
+\sum_{i=1}^N m_i {\ddot \eta}^r_i(t)=\nonumber \\
 &&{\buildrel {def} \over =}\,
 {{\partial [\rho U^r U^s -\sigma^{rs}](t,\vec \sigma )}\over
{\partial \sigma^s}}.
\label{a6}
\end{eqnarray}

For a system of free particles we have ${\ddot {\vec \eta}}_i(t)\,
{\buildrel \circ \over =}\, 0$ so that $\sigma^{rs}(t,\vec \sigma
)=0$. If there are interparticle interactions, they will determine the
effective stress tensor.

Let us consider an arbitrary point $\vec \eta (t)$. The {\it multipole
moments} of the mass density $\rho$ and momentum density $\rho \vec U$
and of the stress-like density $\rho U^rU^s$  with respect to the
point $\vec \eta (t)$ are defined by setting ($N \geq 0$)

\begin{eqnarray}
 m^{r_1...r_n}[\vec \eta (t) ]&=& \int d^3\sigma
[\sigma^{r_1}-\eta^{r_1}(t)]...[\sigma^{r_n}-\eta^{r_n}(t)]\rho (\tau
,\vec \sigma )=\nonumber \\
 &=&\sum_{i=1}^N m_i
[\eta^{r_1}_i(\tau )-\eta^{r_1}(t)]...[\eta^{r_n}_i(t)-\eta^{r_n}(t)],
 \nonumber \\
 &&n=0\quad\quad m[\vec \eta (t)]=m=\sum_{i=1}^N m_i,\nonumber \\
 &&{}\nonumber \\
 p^{r_1...r_nr}[\vec \eta (t) ]&=& \int d^3\sigma
[\sigma^{r_1}-\eta^{r_1}(t)]...[\sigma^{r_n}-\eta^{r_n}(t)]\rho
(t,\vec \sigma )U^r(t,\vec \sigma )=\nonumber \\
 &=&\sum_{i=1}^N m_i {\dot \eta}^r_i(t)
[\eta^{r_1}_i(t)-\eta^{r_1}(t)]...[\eta^{r_n}_i(t)-\eta^{r_n}(t)],
\nonumber \\
 && n=0\quad\quad p^r[\vec \eta (t)]=\sum_{i=1}^Nm_i{\dot \eta}_i(t)
 =\sum_{i=1}^N\kappa_i^r=\kappa^r_{+}\approx 0,\nonumber \\
  &&{}\nonumber \\
  p^{r_1...r_nrs}[\vec \eta (t) ]&=& \int d^3\sigma
[\sigma^{r_1}-\eta^{r_1}(t)]...[\sigma^{r_n}-\eta^{r_n}(t)]\rho
(t,\vec \sigma )U^r(t,\vec \sigma )U^s(t,\vec \sigma )=\nonumber \\
 &=&\sum_{i=1}^N m_i {\dot \eta}^r_i(t){\dot \eta}^s_i(t)
[\eta^{r_1}_i(t)-\eta^{r_1}(t)]...[\eta^{r_n}_i(t)-\eta^{r_n}(t)].
 \label{a7}
 \end{eqnarray}

 The {\it mass monopole} is the conserved mass, while the {\it momentum monopole}
 is the total 3-momentum, which vanishes in the rest frame.

The point $\vec \eta (t)$ is the {\it center of mass} if the {\it mass
dipole} vanishes

\begin{equation}
m^r[\vec \eta (t)]=\sum_{i=1}^Nm_i[\eta^r_i(t)-\eta^r(t)]=0
\Rightarrow \vec \eta (t)={\vec q}_{nr}.
\label{a8}
\end{equation}

The time derivative of the mass dipole is

\begin{equation}
{{d m^r[\vec \eta (t)]}\over {dt}} = p^r[\vec \eta (t)]-m {\dot
\eta}^r(t)=\kappa_{+}^r-m{\dot \eta}^r(t).
\label{a9}
\end{equation}

When $\vec \eta (t)={\vec q}_{nr}$, from the vanishing of this time
derivative we get the {\it momentum-velocity relation for the center
of mass}

\begin{equation}
p^r[{\vec q}_{nr}]=\kappa^r_{+} = m {\dot q}_{+}^r \quad [\approx 0\,
in\, the\, rest\, frame].
\label{a10}
\end{equation}

The {\it mass quadrupole} is

\begin{equation}
m^{rs}[\vec \eta
(t)]=\sum_{i=1}^Nm_i\eta^r_i(t)\eta^s_i(t)-m\eta^r(t)\eta^s(t)-
\Big( \eta^r(t)m^s[\vec \eta (t)]+\eta^s(t)m^r[\vec \eta (t)]\Big),
\label{a11}
\end{equation}

\noindent so that the {\it barycentric mass quadrupole and  tensor of inertia}
are respectively

\begin{eqnarray}
m^{rs}[{\vec q}_{nr}]&=&\sum_{i=1}^N m_i\eta^r_i(t)\eta^s_i(t)-m
q^r_{nr}q^s_{nr},\nonumber \\
 &&{}\nonumber \\
 I^{rs}[{\vec q}_{nr}]&=&\delta^{rs} \sum_um^{uu}[{\vec q}_{nr}]-m^{rs}[{\vec
 q}_{nr}]=\nonumber \\
 &=&\sum_{i=1}m_i[\delta^{rs} {\vec \eta}_i^2(t)-\eta^r_i(t)\eta^s_i(t)]-
 m[\delta^{rs}{\vec q}^2_{nr}-q^r_{nr}q^s_{nr}]=\nonumber \\
 &=&\sum_{a,b}^{1...N-1}k_{ab}({\vec \rho}_a\cdot {\vec
\rho}_b\delta^{rs}-\rho^r_a\rho^s_b) ,\nonumber \\
\Rightarrow&& m^{rs}[{\vec q}_{nr}]=\delta^{rs}\sum_{a,b=1}^{N-1}k_{ab}\,
{\vec \rho}_a\cdot {\vec \rho}_b-I^{rs}[{\vec q}_{nr}].
\label{a12}
\end{eqnarray}

The antisymmetric part of the barycentric momentum dipole gives rise
to the {\it spin vector} in the following way

\begin{eqnarray}
p^{rs}[{\vec q}_{nr}]&=&\sum_{i=1}^Nm_i\eta^r_i(t){\dot
\eta}^s_i(t)-q^r_{nr}p^s[{\vec q}_{nr}]=\sum_{i=1}^N\eta^r_i(t)\kappa_i^s(t)-
q^r_{+}\kappa^s_{+},\nonumber \\
 &&{}\nonumber \\
 S^u&=&{1\over 2}\epsilon^{urs} p^{rs}[{\vec q}_{nr}]=\sum_{a=1}^{N-1}
 ({\vec \rho}_a\times {\vec \pi}_{qa})^u.
\label{a13}
\end{eqnarray}

The {\it multipolar expansions} of the mass and momentum densities
around the point  $\vec \eta (t)$ are

\begin{eqnarray}
\rho (t,\vec \sigma )&=& \sum_{n=0}^{\infty} {{m^{r_1....r_n}[\vec \eta ]}\over {n!}}
{{\partial^n}\over {\partial \sigma^{r_1}...\partial \sigma^{r_n}}}
\delta^3(\vec \sigma -{\vec \eta }(t)),\nonumber \\
&&{}\nonumber \\
\rho (t,\vec \sigma ) U^r(t,\vec \sigma )
 &=&\sum_{n=0}^{\infty}{{p^{r_1....r_nr}[\vec \eta ]}\over {n!}}
{{\partial^n}\over {\partial \sigma^{r_1}...\partial \sigma^{r_n}}}
\delta^3(\vec \sigma -{\vec \eta }(t)).
\label{a14}
\end{eqnarray}

For the {\it barycentric multipolar expansions} we get

\begin{eqnarray}
\rho (t,\vec \sigma )&=&m \delta^3(\vec \sigma -{\vec q}_{nr})-
{1\over 2}(I^{rs}[{\vec q}_{nr}]-{1\over
2}\delta^{rs}\sum_uI^{uu}[{\vec q}_{nr}]){{\partial^2}\over {\partial
\sigma^r\partial \sigma^s}} \delta^3(\vec \sigma -{\vec q}_{nr})+
\nonumber \\
&+&\sum_{n=3}^{\infty} {{m^{r_1....r_n}[{\vec q}_{nr}]}\over {n!}}
{{\partial^n}\over {\partial \sigma^{r_1}...\partial \sigma^{r_n}}}
\delta^3(\vec \sigma -{\vec q}_{nr}),\nonumber \\
&&{}\nonumber \\
\rho (t,\vec \sigma ) U^r(t,\vec \sigma )&=&\kappa^r_{+} \delta^3(\vec \sigma
-{\vec q}_{nr})+\Big[ {1\over 2}\epsilon^{rsu}S^u+p^{(sr)}[{\vec q}_{nr}]
\Big] {{\partial}\over {\partial \sigma^s}} \delta^3(\vec \sigma -
{\vec q}_{nr})+\nonumber \\
 &+&\sum_{n=2}^{\infty}{{p^{r_1....r_nr}[{\vec q}_{nr}]}\over {n!}}
{{\partial^n}\over {\partial \sigma^{r_1}...\partial \sigma^{r_n}}}
\delta^3(\vec \sigma -{\vec q}_{nr}).
\label{a15}
\end{eqnarray}

\vfill\eject

\section{Symmetric Trace-Free Tensors.}

In the applications to gravitational radiation
\cite{sachs,pir,thorne,luc} one does not use {\it Cartesian
tensors} but {\it irreducible symmetric trace-free Cartesian
tensors (STF tensors)}. While a {\it Cartesian multipole tensor of
rank $l$} (like the rest-frame Dixon multipoles) on $R^3$ has
$3^l$ components, of which in general ${1\over 2}(l+1)(l+2)$ are
independent, a {\it spherical multipole moment of order $l$} has
only $2l+1$ independent components. Even if spherical multipole
moments are preferred in calculations of molecular interactions,
spherical harmonics have various disadvantages in numerical
calculations: for analytical and numerical calculations the
Cartesian moments are often more convenient (see for instance
Ref.\cite{hinsen} for the case of the electrostatic potential).
Therefore one prefers to use the {irreducible Cartesian STF
tensors}\cite{gel,coope} (with $2l+1$ independent components if of
rank $l$), which are obtained by using {\it Cartesian spherical
(or solid) harmonic tensors} in place of spherical harmonics.

Given an Euclidean tensor $A_{k_1...k_I}$ on $R^3$, one defines the
completely symmetrized tensor $S_{k_1..k_I} \equiv A_{(k_1..k_I)} =
{1\over {I!}} \sum_{\pi} A_{k_{\pi (1)}...k_{\pi (I)}}$. Then, the
associated STF tensor is obtained by removing all traces ($[I/2]=$
largest integer $\leq I/2$):

\bea
 A_{k_1...k_I}^{(STF)} &=&\sum_{n=0}^{[I/2]} a_n\, \delta_{(k_1k_2} ...
\delta_{k_{2n-1}k_{2n}} S_{k_{2n+1}...k_I) i_1i_1...j_nj_n},\nonumber \\
 &&{}\nonumber \\
 a_n &\equiv& (-1)^n {{ l! (2l-2n-1)!!}\over {(l-2n)! (2l-1)!!(2n)!!}}.
 \label{b1}
 \eea

\noindent For instance $(T_{abc})^{STF} \equiv T_{(abc)}- {1\over 5}\Big[\delta_{ab}
T_{(iic)}+ \delta_{ac}T_{(ibi)}+\delta_{bc}T_{(aii)}\Big]$.

\vfill\eject

\end{document}